\documentstyle[emulapj]{article}
\input epsf.def
\font\tenbg=cmmib10 at 10pt
\def \rvecxi{{\hbox{\tenbg\char'030}}}
\def \rvecphi{{\hbox{\tenbg\char'036}}}
\def \rvecdelta {{\hbox {\tenbg\char'016}}} 
\def \rvecepsilon {{\hbox {\tenbg\char'017}}}

\begin{document}

\title{Eccentric Behavior of Disk Galaxies} 

\author{R.V.E. Lovelace, }
\affil{Los Alamos National Laboratory, 
T-6, MS B288, Los Alamos, NM 87545,
and Department of Astronomy, 
Cornell University, Ithaca, NY 14853;
rvl1@cornell.edu} 

\author{L. Zhang}
\affil{Department of Physics,
Cornell University, Ithaca, NY 14853; lz15@cornell.edu}
\author{ D.A. Kornreich and M.P. Haynes}
\affil{Department of Astronomy,
Cornell Univeristy, Ithaca, NY 14853;
kornreic@astrosun.tn.cornell.edu;
haynes@astrosun.tn.cornell.edu}

\begin{abstract}

  A theory is developed for 
the dynamics of eccentric  perturbations
$[\propto \exp(\pm i\phi)]$
of a disk galaxy residing in a spherical
dark matter halo and including a spherical
bulge component. 
  The disk is represented as a large number $N$
of rings with shifted centers {\it and}
with perturbed azimuthal matter distributions.
Account is taken of the dynamics
of the shift of the matter at
the galaxy's center which may
include a massive black hole.
  The gravitational interactions between the
rings and between the rings
and the center is fully accounted for, but the 
halo and bulge components are treated
as passive gravitational field sources.
  Equations of motion and a Lagrangian
are derived for the ring$+$center system, and these
lead to total energy and total angular
momentum constants of the motion.

  We first study the eccentric motion
of a disk consisting of two rings of
different radii but equal mass, $M_d/2$.
  For  small $M_d$ the two rings 
are stable, but for $M_d$ larger than a
threshold value the rings are unstable
with a dynamical time-scale growth.
  For $M_d$ sufficiently above this threshold,
the instability acts to decrease the
angular momentum of the inner ring, while
increasing that of the outer ring.  
  The instability results from the merging
positive and negative energy modes with 
increasing $M_d$.
  Secondly, we analyze the eccentric motion
of one ring interacting with a radially shifted
central mass.  
    In this case instability sets
in above a threshold value of the central
mass (for a fixed ring mass), and it  
acts to increase the angular momentum of
the central mass (which therefore rotates
in the direction of the disk matter), while
decreasing the angular momentum of the ring.

   Thirdly, we study the eccentric dynamics
of a disk with an exponential surface density
distribution represented by a large number
of rings.  
   The inner part of the disk is
found to be strongly unstable. 
  Angular momentum of the rings is
transferred outward {\it and} to
the central mass if  present, and
a trailing one-armed spiral wave is formed in
the disk.
  Fourthly, we analyze a disk 
with a modified exponential
density distribution where the density of
the inner part of the disk is reduced.
  In this case we find much slower, linear
growth of the eccentric motion. 
  A trailing
one-armed  spiral wave forms in
the disk and  becomes more
tightly wrapped as time increases.
  The motion of the central mass if present 
is small compared with that of the disk.

\end{abstract}

\section{Introduction}

  Although 
studies of spiral galaxies 
commonly assume an axisymmetric
equilibrium state, possibly perturbed
by spiral arms,
there is growing evidence that
many galaxies lack  this
symmetry. 
    Based on optical 
appearance, approximately $30\%$ of disk
galaxies exhibit significant 
``lopsidedness'' (Rix \& Zaritsky 1995;
Kornreich, Haynes, \& Lovelace 1998) which
supports the early findings of Baldwin,
Lynden-Bell, \& Sancisi (1980).
     As many as $\sim50\%$
of spiral galaxies show 
departures from the expected symmetric
two--horned global HI line profile 
(Richter \& Sancisi 1994; Haynes
{\it et al.} 1998). 
  Furthermore, HI maps of several galaxies,
for example, NGC~3631 (Knapen 1997), 
NGC~5474 (Rownd, Dickey, \& Helou 1994), and
NGC~7217 (Buta {\it et al.} 1995), 
have revealed offsets between the
optical centers of light and 
their kinematic centers. 
  Two examples of kinematically
lopsided galaxies 
have recently
be discussed by Swaters {\it et al.} (1998).
    Other recent
observations such as the 
{\it Hubble Space Telescope\/} 
observations
of the nucleus of M31 
(Lauer {\it et al.} 1993) 
and the Kitt Peak 0.9
meter observations of NGC~1073 
(Kornreich {\it et al.} 1998)
indicate that even the optical 
center of light may be displaced from
the center of the optical 
isophotes of the major part of the galaxy in
a significant fraction of cases.

  The origin  of the lopsideness
in disk galaxies remains uncertain. 
  Baldwin 
{\it et al.} (1980) proposed a 
simple kinematic model in which 
different rings of the galaxy,
assumed non-interacting, are initially
shifted from their centered 
equilibrium positions.
  The shifted rings precess in the overall
gravitational potential in a direction
opposite to that of the mass motion.
  Because the precession rate
decreases in general with radial 
distance, an initial disturbance tends
to  ``wind up'' into a leading
spiral arm in a time appreciably
less than the Hubble time. 
   Miller and Smith (1988, 1992) have
made extensive computer simulation
studies of the  unstable
eccentric motion of matter in the
nuclei of galaxies which they
suggest is pertinent to the off-center
nuclei observed in a number of 
galaxies, for example, M31, M33, and M101.

   Understanding of the origin 
of the observed disk asymmetries 
is important because it
provides clues to ongoing 
accretion of gas and the distribution 
of dark, unseen mass in the
halo. 
  Schoenmakers, Franx, and de Zeeuw (1997), 
interpret optical
asymmetry as an indicator of 
asymmetry in the overall galactic
potential, and therefore an 
indicator of the  spatial
distribution of the dark 
matter in a galaxy which may have
a triaxial distribution.
   By analyzing the spiral
components present in the 
surface brightness or \ion{H}{1}
distribution, the velocity 
gradient, and therefore, the shape of the
gravitational potential, may be uncovered.
  Jog (1997)
has studied of the 
orbits of stars and gas in a lopsided
potential, and shows that 
lopsided potentials arising from disks
alone are not self-consistent; 
rather, a stationary lopsided disk may
be responding to asymmetries in the halo.

   Zaritsky and Rix (1997) 
proposed that optical
lopsidedness arises from tidal 
interactions and/or minor mergers. 
  Such mergers are often suggested  as 
the most likely contributors to galaxy
asymmetry, even when no 
interacting companions are evident. 
    While galaxies such as
NGC~5474 are well-known to 
be under the tidal
influence of  neighbors, 
the apparently long-lived kinematic
offsets of other relatively 
isolated objects, and the common
asymmetries in flocculent 
(as opposed to tidally induced ``grand
design'') spiral galaxies, 
are not explained by simple tidal
interaction models, which 
produce only transient asymmetric
features.

   A further possibility is that an optical disk 
may be in a quasi-stationary lopsided
state in a symmetric potential, 
as discussed by Syer and Tremaine (1996). 
  In this model,
gaseous and stellar matter swirl about the
minimum of the halo potential 
in a state not fully relaxed. 
  The result is a lopsided flow within a 
symmetric mass distribution. 
   Numerical simulations of this situation
have been done by Levine and Sparke (1998) using
a gravitational $N-$body 
tree-code method (see Barnes \&
Hut 1989) for  disk galaxies  
shifted from  the center of the main halo
potential. The results are 
suggestive of lopsidedness with large
lifetimes. 
  An $N-$body simulation study 
of a rotating spheroidal
stellar system including the dynamics
of a massive central object by 
Taga and Iye (1998a) indicates that
the central object goes into a
long lasting oscillation similar
to those found earlier by Miller and
Smith (1988, 1992) and
which may explain asymmetric 
structures observed
in M31 and NGC 4486B. 
  A linear stability analysis of
a self-gravitating fluid disk 
including a massive
central object also by Taga and Iye (1998b)
indicates a linear instability 
(Taga \& Iye 1998b).
   We comment on the relation of this
work to the present study in the
conclusions section of this work.

  Here, we develop a theory of
the dynamics of eccentric perturbations
of a disk galaxy residing in a spherical
dark matter halo.  
     We represent
the disk as a large number $N$
of rings as suggested by Baldwin {\it et al.} (1980)
(and Lovelace (1998) for the treatment
of disk warping).  
  In contrast with Baldwin {\it et al.}, the
gravitational interactions between the
rings is fully accounted for.
   We show that for general 
eccentric perturbations,
the centers of the rings are shifted
{\it and} the azimuthal distribution
of matter in the rings is perturbed.
The ring representation is
analogous to the approach of Contopoulos and
Gr{\o}sbal (1986, 1988), where self-consistent
galaxy models are constructed from a finite
set of stellar orbits.

   Section 2 develops a theory
for treating eccentric perturbations
of a disk galaxy.  
   The assumed equilibrium
is first discussed (\S 2.1), and a 
description of the disk perturbations
is developed (\S 2.2).  
   The representation
of the disk in terms of a finite
number $N$ of rings
is presented (\S 2.3), and the ring
equations of motion are derived (\S 2.4).
  We renormalize the ring
equations so as to reduce the nearest
ring interactions (\S 2.5).  
   The dynamics and influence
of the displacement of
the center of the galaxy, which may
include a massive black hole, is 
discussed separately (\S 2.6).
  We obtain an energy constant of the
motion for the dynamical equations (\S 2.7),
the Lagrangian,
 and the
conserved total canonical angular
momentum (\S 2.8).

    We  discuss the nature
of the precession of
a single ring in \S 3.
   In \S 4  we study the eccentric motion
of a disk consisting 
of two rings and  show that this
motion is unstable for sufficiently
large ring masses.
   In \S 5 we study the eccentric motion
of one ring including the radial shift
of the central  mass and show that
this situation is unstable for sufficiently
large mass of  the center and/or of
the ring.
  Section 6 presents numerical results
for the eccentric dynamics of disk
of many rings including the radial shift
of the central mass.
   Section 7 summarizes the conclusions
of this work.

%%%%%%%%%%%%%%%%%%%%%%%%%%%%%%%%%%%%%%%%%%
\begin{figure*}[t]
\epsfscale=500
\plotone{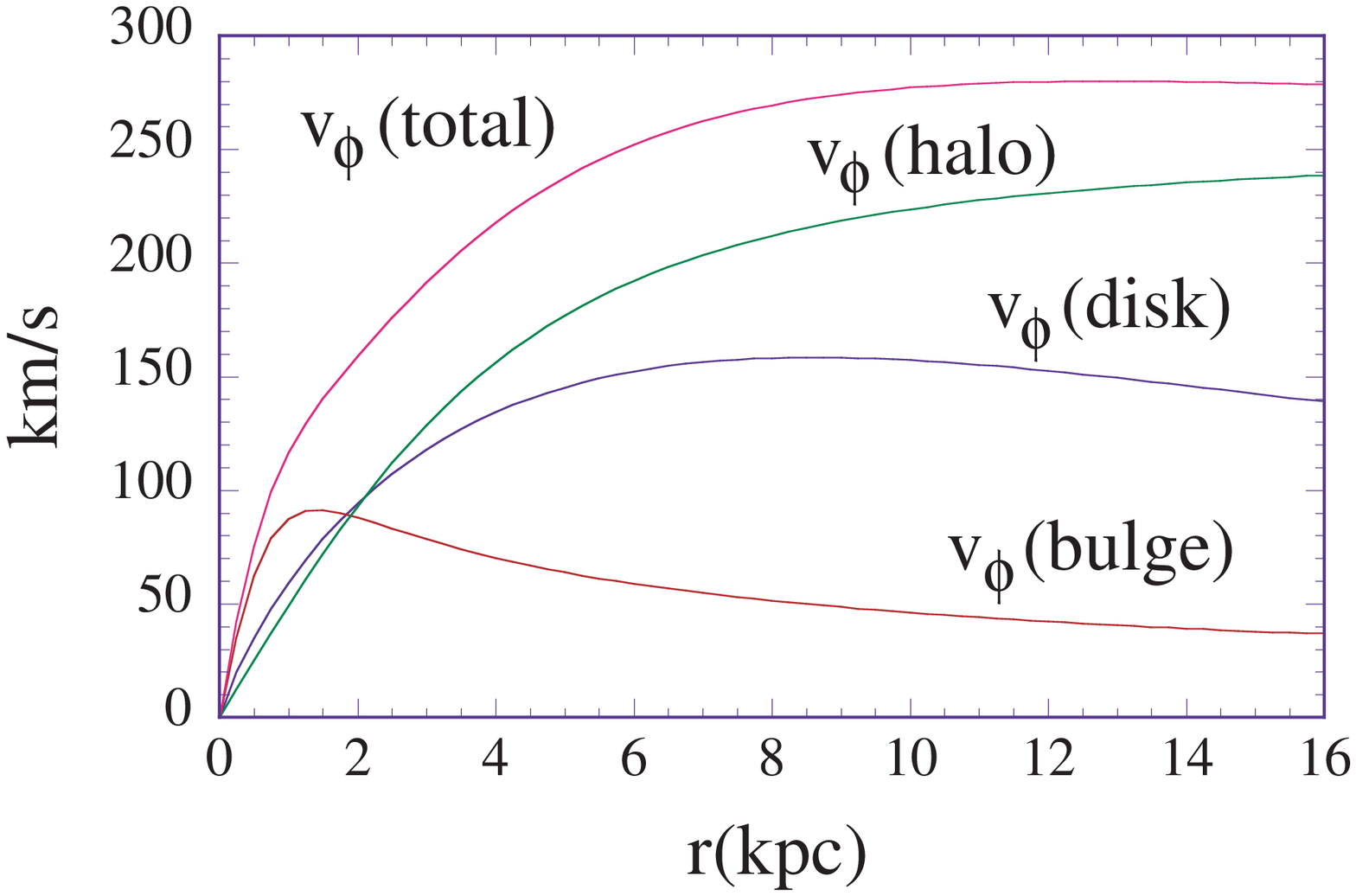}
\caption{
Sample disk rotation curve 
$v_\phi(r)$
for the  values 
$M_d = 6\times 10^{10}M_\odot$ and
$r_d=4$ kpc for the disk,
$M_b=5 \times 10^{9}M_\odot$ and  $r_b=1$ kpc for
the bulge, and $v_h=250$ km/s and
$r_h = 5$ kpc for the halo, using
expressions given
in \S 2.1.   
}
\end{figure*}
%%%%%%%%%%%%%%%%%%%%%%%%%%%%%%%%%%%%%%%%%%

\section{Theory}

\subsection{Equilibrium}

The equilibrium galaxy is assumed to be 
axisymmetric and to consist of a thin disk of
stars and gas and a spheroidal 
distributions consisting of
a bulge component and a halo of
dark matter.  
We use an inertial cylindrical $(r,\phi,z)$
and Cartesian $(x,y,z)$ coordinate systems with the
disk and halo equatorial planes in the $z=0$ plane.  
 The
total gravitational potential is written as
\begin{equation}
\Phi(r,z)=\Phi_{d} +\Phi_b+\Phi_h~, 
\end{equation}
where $\Phi_{d}$ is the potential due to the 
disk, $\Phi_b$ is due to the bulge,
and $\Phi_h$ is that for the
halo.  
  The galaxy may have a central massive
black hole of mass $M_{bh}$ in which
case a term $\Phi_{bh}=-GM_{bh}/\sqrt{r^2+z^2}$ is
added to the right-hand side of (1).
 The particle orbits in the equilibrium disk
are approximately circular with angular
rotation rate $\Omega(r)$, where
\begin{equation}
\Omega^2(r)={1\over r}
{{\partial \Phi}\over{\partial r}}\bigg|_{z=0} =
\Omega_{d}^2+\Omega_b^2+\Omega_h^2~.
\end{equation}
The equilibrium disk velocity is 
${\bf v} = r \Omega(r) \hat{\rvecphi~}.$
 A central black hole is  accounted for
by adding the term $\Omega_{bh}^2=GM_{bh}/r^3$
to the right-hand side of (2).

   The surface mass density of the (optical) disk 
is taken to be $\Sigma_d =\Sigma_{d0} {\rm exp}
(-r/r_d)$ with $\Sigma_{d0}$ and $r_d$  
constants and $M_d=2\pi r_d^2\Sigma_{d0}$ the
total disk mass.  The potential
due
to this disk matter is
$$\Phi_d(r,0)=-~{G M_d \over r_d}
R[I_0(R)K_1(R)
-I_1(R)K_0(R)] ~,
$$
and the corresponding angular velocity is
\begin{equation}
\Omega_d^2={1\over 2}{G M_d \over r_d^3}
\left[I_0(R)K_0(R)
-I_1(R) K_1(R) \right]~,
\end{equation}
where $R\equiv r/(2r_d)$
and the $I's$ and $K's$ are
the usual modified Bessel 
functions (Freeman 1970;  Binney \& Tremaine 1987, p.77).
  Typical values are
$M_d = 6 \times 10^{10}{\rm M}_\odot$ 
and $r_d = 4$ kpc.  
  For these values, $v_d \equiv \sqrt{GM_d/r_d}
\approx 255$ km/s.

  The potential due to the bulge component
is taken as a Plummer model
$$
\Phi_b =-~{G M_b \over (r_b^2+r^2+z^2)^{1/2}}~,
$$
where $M_b$ is the mass of the bulge and $r_b$
is its characteristic radius (Binney \& Tremaine
1987, p.42).  
We have
\begin{equation}
\Omega_b^2 = { G M_b \over
(r_b^2 + r^2)^{3/2}}~.
\end{equation}
Typical values
are $M_b = 10^{10} {\rm M}_{\odot}$ and $r_b =
1$ kpc, and for these values $v_b \equiv \sqrt{GM_b/r_b}
\approx 208~ {\rm km/s}$.

 The  potential of the halo is taken to
be 
$$\Phi_h = {1\over 2}v_{h}^2 ~
\ln(r_{h}^2+{ r}^2+z^2)~,
$$
 where
$v_{h}=$ const is the circular 
velocity at large distances and 
$r_{h}=$ const is the core radius 
of the halo. 
We have
\begin{equation}
\Omega_h^2 = {v_{h}^2 \over r_{h}^2+{ r}^2}~.
\end{equation}
Typical values are $v_{h} \sim 200 -300$ km/s 
and $r_{h}\sim 2-20$ kpc.  Figure 1 shows
an illustrative rotation curve.

\placefigure{fig1}

\subsection{Perturbations}

We treat
the disk as fluid and use a 
Lagrangian representation
for the perturbation
as developed by Frieman and Rotenberg (1960).
 The position vector ${\bf r}$ of a fluid 
element which at $t=0$ was at ${\bf r}_0$ is given
by 
%eqn(4)
\begin{equation}
{\bf r} = {\bf r}_0 +\rvecxi({\bf r}_0,t)~.
\end{equation}
   That is, ${\bf r}_0(t)$ is the unperturbed and 
${\bf r}(t)$ the perturbed orbit of a fluid element.
  This description is applicable to {\it both}
the disk gas and the disk stars which are in
approximately laminar motion with circular
orbits.  The perturbations of the
halo and bulge are assumed negligible compared
with that of the disk.  
  For these approximately spheroidal components,
the particle motion is highly non-laminar with
criss-crossing orbits with the result that
their response ``averages out'' the disk
perturbation.

   Further, the perturbations 
are assumed to consist 
of small in-plane displacements or shifts
of the disk matter,
\begin{equation}
\rvecxi =\xi_r \hat{{\bf r}}+
\xi_\phi\hat{\rvecphi~}~,
\end{equation}
with azimuthal mode number  $m=1$. 
   That is, $\xi_r$ and $\xi_\phi$ have
$\phi$-dependences proportional to
$\exp(i\phi)$.

From equation (6), we have 
${\bf v}({\bf r},t)={\bf v}_0({\bf r}_0,t)+
\partial \rvecxi/\partial t + 
({\bf v} \cdot {\bf \nabla})
\rvecxi$.  
The Eulerian velocity perturbation is
$\delta {\bf v}({\bf r},t) \equiv {\bf v}({\bf r},t)
-{\bf v}_0({\bf r},t)$.  
Therefore,
    %eqn(8)
\begin{equation}
\delta{\bf v}({\bf r},t) ={\partial \rvecxi \over \partial t}
+\left({\bf v} \cdot {\bf \nabla}\right)  \rvecxi 
-\left(\rvecxi \cdot {\bf \nabla} \right) {\bf v}~.
\end{equation}
The components of this equation are
     %eqn(9)
\begin{eqnarray}
\delta v_r &=& {\cal D} ~\xi_r~, \nonumber \\
\delta v_\phi &=& {\cal D} ~\xi_\phi -r \Omega^\prime~ \xi_r~,
\end{eqnarray}
where 
$$
{\cal D} \equiv {\partial \over \partial t} 
+\Omega(r){\partial \over \partial \phi}~,
$$ 
and
$\Omega^\prime \equiv \partial \Omega/\partial r$~.

The main equation of motion is
    % eqn(10)
\begin{equation}
{d~ \delta {\bf v} \over dt} = \delta {\bf F} 
= -{\bf \nabla } \delta \Phi~,
\end{equation}
where $\delta {\bf F}$ is the perturbation
in the gravitational force
(per unit mass), and $\delta \Phi$ is
the perturbation of the gravitational
potential.  
 The pressure force contribution
is small compared to $\delta {\bf F}$ by a factor
$(v_{th}/v_\phi)^2 \ll 1$ and is neglected, where 
$v_{th}$ is the `thermal' spread of the velocities
of the disk matter.  
Also, note that
     %eqn(11)
\begin{equation} 
{d~ \delta{\bf v} \over dt}= 
{\partial ~\delta {\bf v} \over \partial t}
+\left({\bf v}\cdot {\bf \nabla} \right) \delta {\bf v}
+\left(\delta {\bf v} \cdot {\bf \nabla} \right) {\bf v}~.
\end{equation}
The components of (11) give
\begin{eqnarray}
\left({d~ \delta {\bf v} \over dt }\right)_r &=&
{\cal D} \delta v_r -2 \Omega \delta v_\phi~, \nonumber \\
&=& ({\cal D}^2 +2\Omega r \Omega^\prime)~ \xi_r
-2\Omega{\cal D}~\xi_\phi~, \nonumber\\
\left({d~ \delta {\bf v} \over dt} \right)_\phi 
&=& {\cal D} \delta v_\phi 
+(\kappa_r^2/2\Omega) \delta v_r~, \nonumber \\
&=& {\cal D}^2 ~\xi_\phi +2\Omega{\cal D} ~\xi_r~,
\end{eqnarray}
where $\kappa_r^2 \equiv (1/r^3)d(r^4\Omega^2)/dr$ 
is the radial 
epicyclic frequency (squared).

The perturbation of the surface mass
density of the disk obeys
\begin{eqnarray*}
{\partial~\delta \Sigma \over \partial t}
= -{\bf \nabla}\cdot \left(\Sigma ~\delta {\bf v}
+\delta \Sigma ~{\bf v} \right)~, 
\end{eqnarray*}
where $\Sigma(r)$ is the surface density of the
equilibrium disk.   
Because ${\bf \nabla} \cdot
(\Sigma~{\bf v}_0) = 0$, this equation implies
% eqn(12)
\begin{equation}
\delta \Sigma = - {\nabla}\cdot 
\left(\Sigma~\rvecxi\right)~.
\end{equation}
The perturbation of the gravitational potential
is given by
    %eqn(14)
\begin{equation}
\delta \Phi({\bf r},t) = - G \int d^2 r^\prime~ 
{\delta \Sigma({\bf r}^\prime,t)
\over \left| {\bf r} - {\bf r^\prime} \right|}~,
\end{equation}
where the integration is over the 
surface area of the disk.

\subsection{Ring Representation}

 We represent the  disk 
by a finite number $N$ 
of radially shifted plane circular rings. 
The matter distribution around each
ring is also perturbed.  An elliptical
distortion of a ring 
corresponds to an $m=\pm2$ which is
not considered here.
  This description is
{\it general} for small shifts 
$( dr/r)^2 \ll1$, where  
the linearized equations are applicable 
(Lovelace 1998). 
  Of course, the orbit of a single
perturbed particle is not in general
closed in the inertial frame used.  However,
the orbit {\it is} closed in an
appropriately rotating frame, and this
rotation rate is simply the angular 
precession frequency of the ring $\omega$
discussed below (Baldwin {\it et al.} 1980).
  The disk is assumed geometrically thin.
 
   For the equilibrium disk we take
\begin{equation}
\Sigma (r) = \sum_{j=1}^N {M_j 
\over 2\pi r \sqrt{2\pi} \Delta r_j}~ 
\exp \left[-~{(r-r_j)^2 \over 2\Delta r_j^2} \right]~,
\end{equation}
where $M_j$ is the mass of the $j^{th}$ 
ring,  $r_j$ is its radius with 
$0<r_1 < r_2 ~...~ r_N$, and $\Delta r_j \ll r_j$
is its width.  
   The motion of the central part of the
disk ($r<r_1$) is treated separately in \S 2.6 
     
    A physical choice for the rings distribution
will have the ring spacing of the order of 
the disk thickness, 
$r_{j+1}-r_j = {\cal O}( \Delta z)$.
  Further, we assume 
$(r_{j+1}-r_j)^2 \ll r_{j+1}r_j$ in order
to simplify the calculation of the ring 
interaction as discussed in Appendix A.
    For example, a possible choice is
$r_j=r_1+(j-1)\delta r$ with
$r_1=1$ kpc, $\delta r=0.5$ kpc,
 and $M_j=2\pi r_j\delta r
\Sigma_d(r_j)$.  
For  $\Delta r_j = 
\delta r/\sqrt{2\ln2}\approx \delta r/1.177$,
the profile of a ring falls to half its
maximum value at $\delta r$ so that
equation (15) 
gives a fairly smooth 
representation of $\Sigma_d(r)$.

 The perturbation in the disk's
surface density is
% equation (16)
\begin{equation}
\delta \Sigma(r,\phi,t) = \sum \delta \Sigma_j~,
\quad
\delta \Sigma_j= 
-{\bf \nabla}\cdot (\Sigma_j \rvecxi)~,
\end{equation}
where $\Sigma_j \!\equiv \!
(M_j/2\pi r \sqrt{2\pi} \Delta r_j)
\exp[- (r-r_j)^2/(2\Delta r_j^2)]$.

We can express different moments of the
perturbed disk in terms of the rings.  
For example, the center of mass of the 
disk is
\begin{equation}
<{\bf r}>
={\sum M_j < {\bf r}_j >
\over \sum M_j}~,
\end{equation}
where 
\begin{eqnarray}
M_j<{\bf r}_j>& =& \int d^2x~ {\bf r} ~\delta \Sigma_j~, 
\nonumber \\
&=& M_j \oint {d\phi \over 2\pi}~  \rvecxi_j~,
\end{eqnarray}
where an integration by parts has been made.

We can write in general
  %equation (19)
\begin{eqnarray}
\xi_{jr} &=& \epsilon_{jx}\cos\phi +
\epsilon_{jy}\sin\phi~,\nonumber \\
\xi_{j\phi}&=& -\delta_{jx}\sin\phi +
\delta_{jy}\cos\phi~.
\end{eqnarray}
Here, $\epsilon_{j x,y}$ and $\delta_{jx,y}$ are
the {\it ring displacement amplitudes}: 
 $\epsilon_{jx,y}$ 
represents the shift of the ring's center, and
$\delta_{jx,y}$ represents in general both
the shift of the ring's center {\it and} the
azimuthal displacement of the ring matter. 
Firstly, notice that for $\epsilon_{jx,y}=0$ and
$\delta_{jx,y} \neq 0$, there is no shift of
the ring's center but rather an azimuthal 
displacement of the ring matter.  In this
case $\delta \Sigma_j = 
-(\Sigma_j/r_j)(\partial \xi_{j\phi}/\partial \phi)$.
  Secondly, notice that $\delta_{jx,y} = \epsilon_{jx,y}$
corresponds to a {\it rigid} shift of the
ring without azimuthal displacement of
the ring matter.  For example, a rigid shift
in the $x-$direction has $\epsilon_{jx}=\delta_{jx}$
and $\epsilon_{jy}=0=\delta_{jy}$ so that
$\xi_{jr}=\epsilon_{jx}\cos\phi$ and $\xi_{j\phi}=
-\epsilon_{jx}\sin\phi$.
 In this case, ${\bf \nabla \cdot}
\rvecxi_j=0$ so that $\delta \Sigma_j = 
-{\bf \nabla \cdot}
(\Sigma_j \rvecxi_j) = 
- \xi_{rj}(\partial \Sigma_j /\partial r)$.
 Figure 2 shows the nature of 
perturbations with a shift of the ring's center
and with an azimuthal displacement of
the ring matter.

\placefigure{fig2}

%%%%%%%%%%%%%%%%%%%%%%%%%%%%%%%%%%%%%%%%%%
\begin{figure*}[t]
\epsfscale=500
\plotone{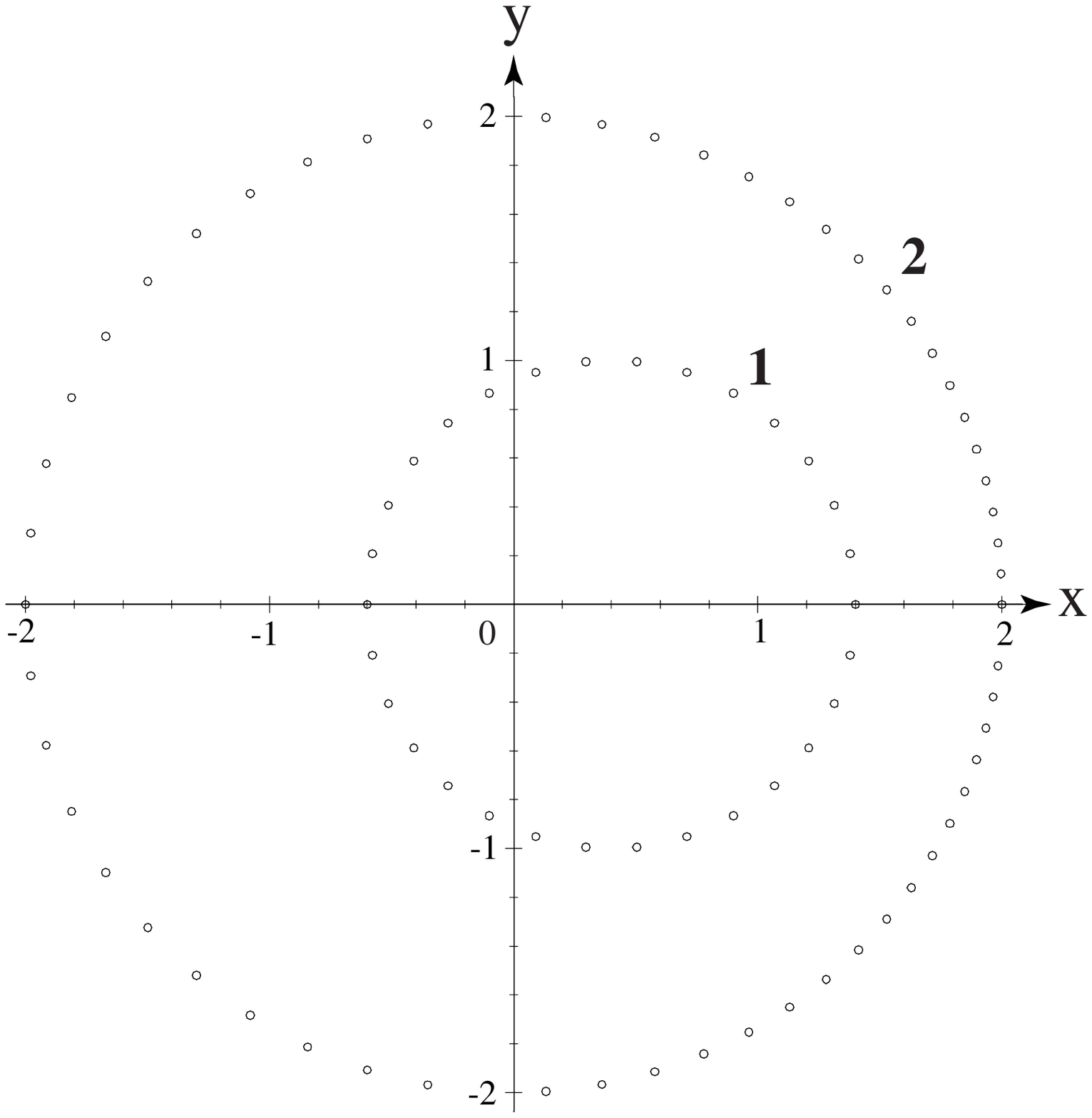}
\caption{
Drawing of two perturbed
rings with
equilibrium radii $r_1=1$ and $r_2=2$ in
arbitrary units.  
 The
center of ring $1$ is rigidly shifted
in the $x-$direction from the origin by 
$\epsilon_{1x}=0.4$ and $\epsilon_{1y}=0$.
The center of ring $2$ is at the origin while
the distribution of matter
around the ring is perturbed as
indicated by the small circles.  For
this ring, $\delta_{2x}=0.4$  and $\delta_{2y}=0$.
}
\end{figure*}
%%%%%%%%%%%%%%%%%%%%%%%%%%%%%%%%%%%%%%%%%%

 Equation (18) now gives
\begin{eqnarray}
<r_{jx}>& =& {1\over 2}(\epsilon_{jx} + \delta_{jx})~,
\nonumber \\
<r_{jy}> &=& {1\over 2}(\epsilon_{jy} + \delta_{jy})~.
\end{eqnarray}
For the case of a rigid shift of a ring,
$\epsilon_{jx,y}=\delta_{jx,y}$, the center of
mass position is simply $<r_{jx,y}>=\epsilon_{jx,y}$,
as expected.

Similarly, the velocity perturbation of
the disk can be written as
\begin{equation}
<\delta {\bf v}>
={\sum M_j < \delta {\bf v}_j >
\over \sum M_j}~,
\end{equation}
where
\begin{equation}
<\delta {\bf v}_j > = {1\over M_j} \int d^2x~
\left(\delta \Sigma_j {\bf v} + 
\Sigma_j \delta {\bf v}_j \right)~.
\end{equation}
Evaluation of (22) gives
\begin{eqnarray}
<\delta v_{jx}>\!&=&\!{1\over2}\left(
\dot{\epsilon}_{jx} 
+\dot{\delta}_{jx} \right),\nonumber\\
<\delta v_{jy}>\!&=&\!{1\over2}\left(
\dot{\epsilon}_{jy}
+\dot{\delta}_{jy} \right)~.
\end{eqnarray}
The influence of the eccentric motion
on the rotation curves is discussed at
the end of \S 6.3.

\subsection{Ring Equations of Motion}

Equations (12) and (19) give the ring
equations of motion,
\begin{eqnarray}
\ddot{\epsilon}_x+2\Omega\dot{\epsilon}_y
-\tilde{\Omega}^2\epsilon_x
-2\Omega (\dot{\delta}_y -\Omega \delta_x)& =&
<\delta F_r^C>~,
\nonumber \\
\ddot{\epsilon}_y-2\Omega\dot{\epsilon}_x
-\tilde{\Omega}^2\epsilon_y
+2\Omega (\dot{\delta}_x +\Omega \delta_y)& =&
<\delta F_r^S>~, 
\nonumber \\
\ddot{\delta}_x+2\Omega\dot{\delta}_y -\Omega^2\delta_x
-2\Omega(\dot{\epsilon}_y -\Omega {\epsilon}_x)& =&
\!\!\!-<\delta F_\phi^S>~, 
\nonumber \\
\ddot{\delta}_y-2\Omega\dot{\delta}_x -\Omega^2\delta_y
+2\Omega(\dot{\epsilon}_x +\Omega {\epsilon}_y) &=&
<\delta F_\phi^C>~, 
\nonumber\\ 
\end{eqnarray}
where the $j$ subscripts are implicit,
where the angular
brackets indicate the
average over the ring
 $<(..)>\equiv 2\pi \int r dr (..) \Sigma_j(r)/M_j$,
where
$\tilde{\Omega}^2 \equiv 
\Omega^2 -2\Omega r\Omega^\prime$, and 
$$
\delta F_\alpha^{C,S} 
\equiv \oint {d\phi \over \pi}~
\left[\cos \phi,~\sin \phi\right]\delta F_\alpha
~,
$$
with $\alpha = r,~\phi$.

We now evaluate the different force terms on
the right-hand side of (24).  For this
it is useful to write $\delta \Sigma =
\delta \Sigma_a + \delta \Sigma_b$, where
\begin{equation}
\delta \Sigma_a =-{1\over r}{\partial (r\Sigma \xi_r)
\over \partial r}~,\quad \delta \Sigma_b =
-{1\over r}{\partial (\Sigma \xi_\phi) \over
\partial \phi}~,
\end{equation}
from equation (13).
  The corresponding contributions to the potential,
$\delta \Phi = \delta \Phi_a + \delta \Phi_b$
evaluated at $(r,\phi)$ 
are from (13),
$$
\delta \Phi_a(r,\phi) =
 G\sum_k M_k(\epsilon_{kx} \cos \phi
+\epsilon_{ky} \sin \phi)~ \times \quad \quad \quad
$$
\begin{equation}
\int_0^\infty r^\prime dr^\prime~
\bigg\{{{1 \over r^\prime}
{\partial \over \partial r^\prime}
\bigg[r^\prime S(r^\prime|r_k)\bigg]}\bigg\}
\oint{d \Psi^\prime \over 2\pi}~
{\cos \Psi \over R(r,r^\prime)}
~,
\end{equation}
and 
$$
\delta \Phi_b(r,\phi) =
- G\sum_k M_k(\delta_{kx}\cos\phi
+\delta_{ky}\sin\phi)~\times \quad \quad \quad
$$
\begin{equation}
\int_0^\infty r^\prime dr^\prime~
{S(r^\prime|r_k) \over r^\prime}
\oint {d\Psi \over 2\pi}~
{\cos \Psi \over  R(r,r^\prime)}
\end{equation}
where $R^2(r,r^\prime) \equiv
r^2+(r^\prime)^2-2rr^\prime \cos\Psi$,
$$
S(r|r_k) \equiv { 2\pi \Sigma_k(r) \over M_k}=
{\exp[-(r-r_k)^2/2\Delta r_k^2]
\over r \sqrt{2\pi} \Delta r_k}~,
$$
and
$\Psi \equiv \phi^\prime -\phi$. 

Evaluation of the force on 
the $j^{\rm th}$ ring due to the other rings gives
%equation (28)
\begin{eqnarray}
M_j <\delta F_{rj}^C> &=& 
\sum_{k}
\left(C_{jk} \epsilon_{kx}
+D_{jk} \delta_{kx}\right)~, \nonumber \\
M_j <\delta F_{rj}^S> &=& 
\sum_{k}
\left( C_{jk} \epsilon_{ky}
+D_{jk}\delta_{ky}\right)~, \nonumber \\
-M_j<\delta F_{\phi j}^S> &=& 
\sum_{k}
\left(E_{jk} \delta_{kx}
+D^\prime_{jk} \epsilon_{kx}\right)~,\nonumber \\
M_j<\delta F_{\phi j}^C> &=& 
\sum_{k}
\left(E_{jk} \delta_{ky}
+D^\prime_{jk} \epsilon_{ky}\right),
\end{eqnarray}
where the `tidal coefficients' are
$$
C_{jk}=-GM_jM_k
\int \!\!\int r dr ~r^\prime dr^\prime ~\times
\quad\quad\quad\quad\quad\quad~~
$$
\begin{equation}
S(r|r_j)
{\partial \big[r^\prime S(r^\prime|r_k)\big]
\over r^\prime ~\partial r^\prime}
{\partial {\cal K}(r,r^\prime) \over \partial r}
~,
\end{equation}
%equation29
$$
D_{jk}=GM_jM_k\int\!\!\int 
r dr~r^\prime dr^\prime~ \times 
\quad\quad\quad\quad\quad\quad\quad\quad
$$
\begin{equation}
{S(r|r_j)}
{S(r^\prime|r_k)} 
{\partial {\cal K}(r,r^\prime) \over r^\prime~\partial r}
\end{equation}
% eqn30
$$
D_{jk}^\prime=-GM_jM_k 
\int\!\!\int rdr~ r^\prime dr^\prime ~\times
\quad\quad\quad\quad\quad\quad\quad
$$
\begin{equation}
{S(r|r_j)\over r}
{\partial \big[r^\prime S(r^\prime|r_k)\big]
\over r^\prime \partial r^\prime}~
{\cal K}(r,r^\prime)
~,
\end{equation}
$$
E_{jk}=GM_jM_k\int\!\!\int rdr~
r^\prime dr^\prime ~\times
\quad\quad\quad\quad\quad\quad\quad\quad
$$
\begin{equation}
{S(r|r_j) }
{S(r^\prime|r_k) }~
{{\cal K}(r,r^\prime) \over r~r^\prime }~,
\end{equation}
where
\begin{equation}
{\cal K}(r,r^\prime) \equiv \oint
{d\Psi \over 2 \pi}
{\cos\Psi \over R(r,r^\prime)}~,
\end{equation}
and where the $r,r^\prime$ integrals
are all from $0$ to $\infty$.

Formal integration by parts of (29) gives
$$
C_{jk}=GM_jM_k
\int \!\!\int r dr ~r^\prime dr^\prime ~\times
\quad\quad\quad\quad\quad\quad~~
$$
\begin{equation}
S(r|r_j)
S(r^\prime|r_k)
{\partial^2 {\cal K}(r,r^\prime) 
\over \partial r^\prime ~\partial r}
~,
\end{equation}
%eqn32
which shows that $C_{jk}=C_{kj}$.
Also, integration by parts of
(31) gives
$$
D_{jk}^\prime=GM_jM_k 
\int\!\!\int rdr~ r^\prime dr^\prime ~\times
\quad\quad\quad\quad\quad\quad\quad
$$
\begin{equation}
{S(r|r_j)}
{ S(r^\prime|r_k)}~
{\partial {\cal K}(r,r^\prime) 
\over r~ \partial r^\prime}
~,
\end{equation}
and this shows that 
$D_{jk} = D^\prime_{kj}$.
Expressions for the tidal coefficients
in terms of elliptic
integrals are given in the Appendix.

Following the approach of Lovelace (1998),
we introduce the complex displacement
amplitudes
\begin{equation}
{\cal E}_j \equiv \epsilon_{jx} - i \epsilon_{jy}
=\epsilon_j(t) \exp[- i\varphi_j(t)]~, 
\end{equation}
\begin{equation}
~~~~~\Delta_j \equiv \delta_{jx}- i \delta_{jy}
= \delta_j(t) \exp[-i \psi_j(t)]~.
\end{equation}
Here, $\epsilon_j \geq 0$ is the amplitude
of the shift of the ring's center, and 
$\varphi_j$ is angle of the shift with
respect to the $x-$axis;  $\delta_j \geq 0$
is the amplitude of the azimuthal displacement
of the ring matter, and $\psi_j$ is the angle
of the maximum of the ring density also
with respect to the $x-$axis.  If $\varphi(t)$
and $\psi(t)$ increase with time, the ring
precesses in the same sense as the particle
motion and we refer to this as {\it forward
precession}.  The opposite case, with $\varphi(t)$
and $\psi(t)$ decreasing with time, is termed
{\it backward precession}.

We now combine equations (24) and (28) to
obtain the ring equations of motion,
% equation (38)
\begin{eqnarray}
M_j \left( ~\ddot{\cal E}_j +
2i\Omega_j \dot{\cal E}_j -
\tilde{\Omega}^2_j {\cal E}_j -
2i\Omega_j \dot{\Delta}_j +2\Omega_j^2 \Delta_j
 \right) = \nonumber \\
\sum_{k}
\left( C_{jk} {\cal E}_k
+D_{jk} \Delta_k \right)~, \\
M_j \left( \ddot{\Delta}_j +
2i\Omega_j \dot{\Delta}_j -
\Omega_j^2 \Delta_j -
2i\Omega_j \dot{\cal E}_j    
+2\Omega_j^2 {\cal E}_j\right)= \nonumber \\
\sum_{k }
\left(E_{jk} \Delta_k+
D^\prime_{jk} {\cal E}_k \right)~,
\end{eqnarray}
where $j=1..N$, $\Omega_j \equiv \Omega(r_j)$, and
$\tilde{\Omega}_j \equiv \tilde{\Omega}(r_j)$.

\subsection{Renormalization of Ring Equations}

Here, we redo the ring equations of
motion so as to diminish the strong
tidal interactions of nearest neighbor
rings due to the terms $\propto C_{j,j+k}$.
First, 
we rewrite the right-hand
side of (38) as
\begin{eqnarray}
\sum_{k}
\left[C_{jk} \left({\cal E}_k-{\cal E}_j\right)
+D_{jk} \left({\Delta}_k-{\cal E}_j \right)\right]+\nonumber\\ 
{\cal E}_j\sum_{k}(C_{jk} + D_{jk})~,
\end{eqnarray}
Similarly, we rewrite the right-hand side of (39)
as
\begin{eqnarray}
\sum_{k}
\left[E_{jk} \left({\Delta}_k-{\Delta}_j\right)
+D_{jk}^\prime \left({\cal E}_k-
{\Delta}_j \right)\right]+\nonumber\\ 
{\Delta}_j\sum_{k}\left( E_{jk} + D_{jk}^\prime \right)~.
\end{eqnarray}

 We define
\begin{equation}
\Omega_{{\cal E}j}^2  \equiv \tilde{\Omega}_j^2+ 
{1\over M_j}\sum_{k}
\left(C_{jk}+D_{jk} \right)~.
\end{equation}
Using the relations $\tilde{\Omega}^2=\Omega^2
-r(d\Omega^2/dr)$,  
$\Omega^2=\Omega_d^2 +\Omega_b^2+\Omega_h^2$, 
and equation (A14)
of the Appendix gives
\begin{equation}
\Omega_{{\cal E}j}^2 =
\Omega_j^2
-\bigg(r{\partial (\Omega_{b}^2+\Omega_h^2)
\over \partial r}\bigg)_j
+{1 \over M_j } 
\sum_{k }(D^\prime_{jk} +E_{jk})~.
\end{equation}
  (If there is a central black hole $M_{bh}$ then
the right-hand side of (43) also has the
term $-[r(\partial \Omega_{bh}^2/\partial r)]_j
=3GM_{bh}/r_j^3$.)
Similarly, we define
\begin{equation}
\Omega_{\Delta j}^2 \equiv \Omega_j^2 +
{1 \over M_j}\sum_{k }(D^\prime_{jk}+E_{jk})~.
\end{equation}
The ring equations of motion now become
%%%%%%%%%%%%%% equation (45/46)
$$
M_j \left( ~\ddot{\cal E}_j +
2i\Omega_j \dot{\cal E}_j -
\Omega_{{\cal E}j}^2 {\cal E}_j -
2i\Omega_j \dot{\Delta}_j +2\Omega_j^2 \Delta_j
 \right) = 
$$
\vspace{-0.12in}
\begin{equation}
\sum_{k}
\left[ C_{jk} \left({\cal E}_k-{\cal E}_j \right)
+D_{jk} \left(\Delta_k-{\cal E}_j\right) \right]~, 
\end{equation}
$$
M_j \left( \ddot{\Delta}_j +
2i\Omega_j \dot{\Delta}_j -
\Omega_{\Delta j}^2 \Delta_j -
2i\Omega_j \dot{\cal E}_j    
+2\Omega_j^2 {\cal E}_j\right)= 
$$
\begin{equation}
\sum_{k }
\left[E_{jk} \left(\Delta_k-\Delta_j\right)+
D^\prime_{jk} \left({\cal E}_k-\Delta_j\right) \right]~.
\end{equation}
In the large $N$ limit,
the sums in equations (45) and (46)
go over to bounded integrals.  
  The
diagonal elements,  $C_{jj}$
(and $E_{jj}$), are absent from (45) and (46).
   Note also that the self-interaction of
a ring vanishes as it should for the case of a rigid 
shift  where ${\cal E}_j = \Delta_j$.

\subsection{Dynamics and Influence of ``Center''}

  The dynamical equations (45) and (46) do
not account for the part of the disk 
inside the innermost ring $r_1$.  Also,
there may be a massive black 
hole $M_{bh}$ near
the galaxy center.  
  We treat this central region
separately as a point
mass $M_0$,  
\begin{equation}
M_0=M_{bh}+2\pi \int_0^{r_1^\prime}
rdr~\Sigma_d(r)~,
\end{equation}
where $r_1^\prime \equiv r_1-\delta r_1/2$.

  The
horizontal displacement of the
``center'' is
% equation 45
\begin{equation}
\rvecepsilon_0(t)=\epsilon_{0x}\hat{\bf x} 
+\epsilon_{0y}\hat{\bf y}~.
\end{equation}
The equation of motion for $\rvecepsilon_0$
taking into account the eccentric displacemnts
of the rings 
is 
%equation 49
\begin{equation}
M_0\bigg( { d^2 {\cal E}_0 \over dt^2} +
\Omega_0^2 {\cal E}_0\bigg) 
=-\sum_{k=1}^N F_{0k}\left( {\cal E}_k - 
{1\over 2} \Delta_k \right)~,
\end{equation}
where 
\begin{equation}
{\cal E}_0 \equiv \epsilon_{0x} - i\epsilon_{0y}
=\epsilon_0\exp(-i\varphi_0)
\end{equation}
is the complex displacement amplitude of the ``center'';
$F_{0k} \equiv GM_0M_k/r_k^3$, $k=1..N$, are
the tidal coefficients between the ``center'' and the
rings; and
\begin{equation}
\Omega_0^2 = 
\bigg({1\over r}{\partial \Phi \over \partial r}\bigg)_0
=\bigg({1\over r}{\partial (\Phi_b+\Phi_h) 
\over \partial r}\bigg)_0
-{1\over 2 M_0}\sum_j F_{0j}
\end{equation}
is the angular oscillation frequency of a particle
at the galaxy center.  
   From \S 2.1, we
have $[(1/r)(\partial \Phi_b/\partial r)]_0 
= GM_b/r_b^3$ and 
$[(1/r)(\partial \Phi_h/\partial r)]_0 
= v_{h}^2/r_{h}^2$.
  For all of the considered conditions,  
we find $\Omega_0^2>0$. 

    The influence of the displaced ``center'' on
the rings is included by adding the force
terms, due to the ``center's''
displacement, to the right-hand sides of equations
(45) and (46),
\begin{equation}
M_j \left( ~\ddot{\cal E}_j ~+~..\right)
=-~2F_{0j}{\cal E}_0~+~..~,
\end{equation}
\begin{equation}
M_j\left(~\ddot{\Delta}_j~+~.. \right) 
=+~F_{0j}{\cal E}_0~+~..~,
\end{equation}
where the ellipses denote terms
in equations (45) and (46).

\subsection{Energy Conservation}

An energy constant of the motion of the ring
system can be obtained by multiplying
equation (52) by $\dot{\cal E}_j^*$
and (53) by $\dot{\Delta}_j^*$ 
(with $[...]^*$ denoting
the complex conjugate), adding
the two equations, summing over
$j$, and dividing by 2.  
(This factor of $2$ makes the
kinetic energy of a rigidly
shifted ring with $\rvecepsilon_j=
\rvecdelta_j$ equal to 
$(1/2)M_j \dot{\rvecepsilon}_j^2$.)
 Further, we multiply equation (49)
by ${\cal E}_0^*$ and add the result to
the previous sum.  
In this way we find $d E/dt =0$, where
\vspace{-0.05in}
$$
E\!=\!{1\over 4}\sum_j M_j
\left(\dot{\rvecepsilon}_j^2 -
{\Omega}^2_{{\cal E} j}\rvecepsilon^2_j +
\dot{\rvecdelta}_j^2 -
\Omega_{\Delta j}^2 \rvecdelta_j^2 +
4\Omega_j^2 \rvecepsilon_j \cdot \rvecdelta_j
\right)
$$
\vspace{-0.1in}
$$
 +{1\over 8}\sum_{k\neq j}\sum_{j}
\bigg[
C_{jk} (\rvecepsilon_j - \rvecepsilon_k)^2
+2D_{jk}(\rvecepsilon_j- \rvecdelta_k)^2+
$$
\vspace{-0.1in}
$$
+E_{jk}(\rvecdelta_j - \rvecdelta_k)^2 \bigg]
$$
\begin{equation}
+{1\over 2}M_0 (\dot{\rvecepsilon}_0^2
+\Omega_0^2 {\rvecepsilon}_0^2)
+\sum_jF_{0j}\left(\rvecepsilon_0\cdot \rvecepsilon_j
-{1\over 2}\rvecepsilon_0\cdot \rvecdelta_j \right)~,
\end{equation}
and where the real vectors  
$\rvecepsilon=\epsilon_x\hat{\bf x}
+\epsilon_y \hat{\bf y}$ and 
$\rvecdelta=\delta_x\hat{\bf x}
+\delta_y \hat{\bf y}$ are useful here.

\subsection{Lagrangian}

By inspection, we find the Lagrangian
for the ring system ${\cal L}
(\epsilon_{jx},\dot{\epsilon}_{jx}, ..)$,
\vspace{-0.05in}
$$
{\cal L}={1\over 4}\sum_j M_j
\bigg\{ \dot{\rvecepsilon}_j^2 +
{\Omega}^2_{{\cal E}j}\rvecepsilon^2_j +
\dot{\rvecdelta}_j^2 +
\Omega_{\Delta j}^2 \rvecdelta_j^2 -
4\Omega_j^2 \rvecepsilon_j\cdot \rvecdelta_j
$$
\vspace{-0.15in}
$$
-2\Omega_j [(\rvecepsilon_j-\rvecdelta_j)
\times (\dot{\rvecepsilon}_j -\dot{\rvecdelta}_j)
 ]\cdot \hat{\bf z} \bigg \}+ 
{1\over 2}M_0 
(\dot{\rvecepsilon}_0^2-\Omega_0^2\rvecepsilon_0^2)
$$
$$
- {1\over 8}\sum_{k\neq j}\sum_{j}
\bigg[
C_{jk} (\rvecepsilon_j - \rvecepsilon_k)^2
+2D_{jk}(\rvecepsilon_j- \rvecdelta_k)^2+
$$
\begin{equation}
\vspace{-0.1in}
+E_{jk}(\rvecdelta_j - \rvecdelta_k)^2 \bigg]
-\sum_j F_{0j}\left(\rvecepsilon_0\cdot \rvecepsilon_j
-{1\over 2}\rvecepsilon_0\cdot \rvecdelta_j \right)~,
\end{equation}

\noindent Because $\partial {\cal L}/\partial t = 0$,
the Hamiltonian,
\begin{equation}
{\cal H} \equiv \sum_j\left(\dot{\epsilon}_{jx}
{\partial {\cal L} \over 
\partial \dot{\epsilon}_{jx}} + ... \right)
-{\cal L}~,
\end{equation}
is a constant of the motion.
It is readily verified that ${\cal H}=E$.

We can make a canonical transformation
$(\epsilon_{jx},\epsilon_{jy})$  $ \rightarrow
(\epsilon_j,\varphi_j)$,
$(\delta_{jx},\delta_{jy}) \rightarrow
(\delta_j,\psi_j)$ to obtain
the Lagrangian as ${\cal L}=
{\cal L}(\dot{\epsilon}_j,\dot{\delta}_j,
\epsilon_j, \delta_j, \varphi_j,\psi_j)$.
Note for example that $\dot{\rvecepsilon}^2_j
\rightarrow \dot{\epsilon}_j^2 + \epsilon_j^2
\dot{\varphi}_j^2$.
It is then clear from the azimuthal
symmetry of the equilibrium that ${\cal L}$ is
invariant under the simultaneous changes
$\varphi_j \rightarrow \varphi_j+\theta$,
$\psi_j \rightarrow \psi_j+\theta$ for
$j=1,..,N$, where $\theta$ is an arbitrary
angle.  Thus
$$
\sum_j\left({\partial {\cal L} 
\over \partial \varphi_j} +
{\partial {\cal L} \over \partial \psi_j} \right)
=0~,
$$
and consequently the total canonical angular
momentum of the ring system,
\begin{equation}
{\cal P}_\phi \equiv  \sum_j
\left({\partial {\cal L} \over 
\partial \dot{\varphi}_j} +
{\partial {\cal L} \over 
\partial \dot{\psi}_j } \right)~,
\end{equation}
is another constant of the motion.
Evaluating (57) gives
\vspace{-0.05in}
$$
{\cal P}_\phi = {1\over 2}\sum_j M_j
\bigg[\epsilon_j^2(\dot{\varphi}_j-\Omega_j)
+\delta_j^2(\dot{\psi}_j-\Omega_j)
$$
\vspace{-0.1in}
\begin{equation}
+2\Omega_j\epsilon_j\delta_j\cos(\varphi_j-\psi_j)
\bigg]+M_0\epsilon_0^2\dot{\varphi}_0~,
\end{equation}
where the last term represents the angular momentum
of the galaxy center. 
   The last term within the square
brackets can also be written 
as $2\Omega_j (\rvecepsilon_j
\cdot \rvecdelta_j)$. 
   
  The constants of the motion ${\cal H}$ and
${\cal P}_\phi$ are valuable for checking
numerical integrations of the equations
of motion (52) and (53).

%%%%%%%%%%%%%%%%%%%%%%%%%%%%%%%%%%%%%%%%%%
\begin{figure*}[t]
\epsfscale=400
\plotone{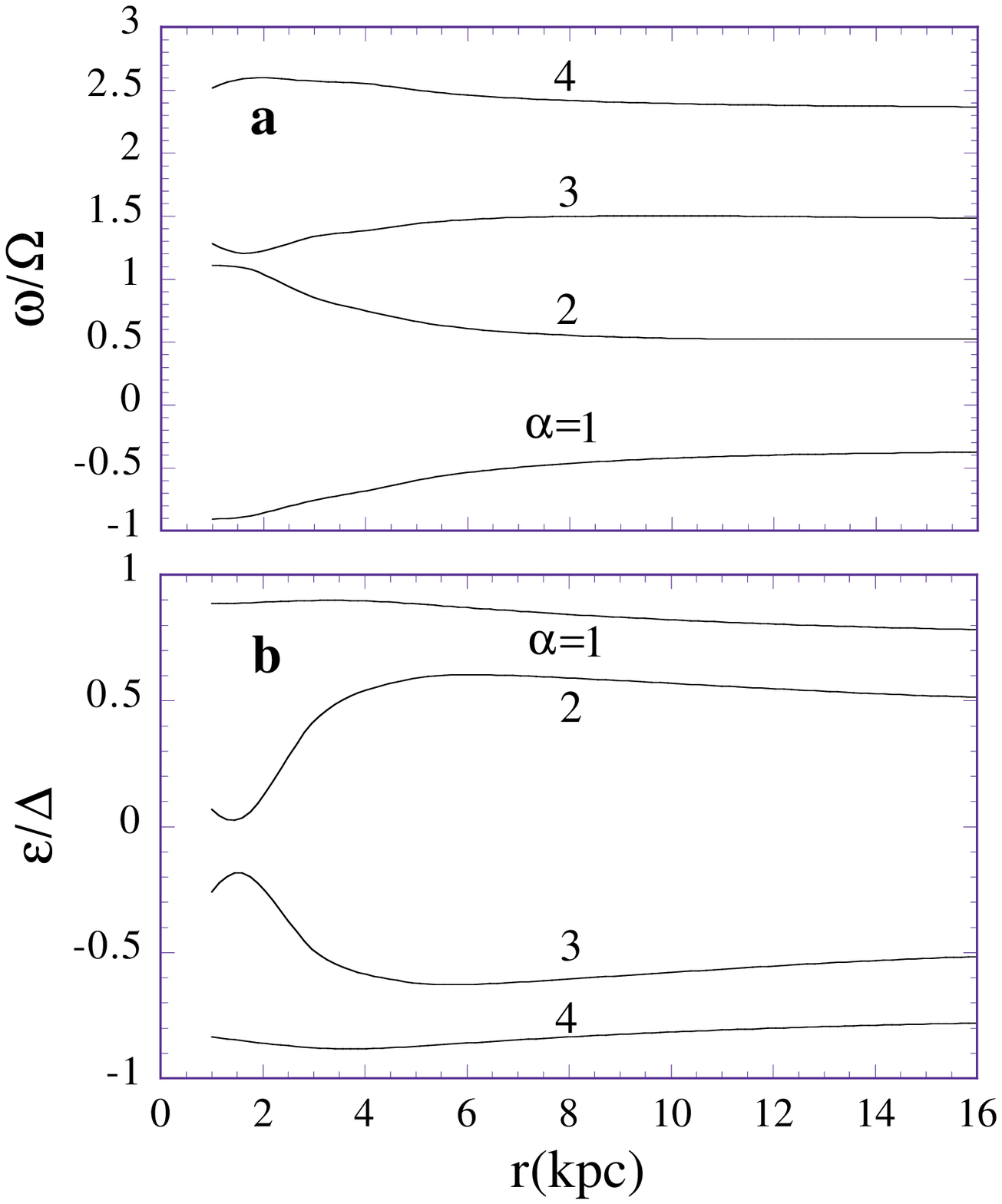}
\caption{
The top panel ({\bf a}) of
the figure shows the radial
dependence of the four frequencies
$\omega_\alpha$ 
($\alpha=1,..,4$) of the modes of
oscillation of an eccentric
ring normalized by the disk's angular
rotation rate $\Omega(r)$. 
  The bottom panel ({\bf b}) shows the
corresponding radial dependence
of the mode amplitude 
ratio ${\cal E}/\Delta$
as discussed in
the text.  
  The case shown is
for the galaxy parameters of
Figure 1 with innermost
ring of radius $r_1=1$ kpc.
  The values
$\Omega_j$, $\Omega_{{\cal E}j}$,
and $\Omega_{\Delta j}$ were
obtained using equations (A11),
(43), and (44).
}
\end{figure*}
%%%%%%%%%%%%%%%%%%%%%%%%%%%%%%%%%%%%%%%%%%

\section{Eccentric Motion of a Single Ring}

   Consider the 
eccentric motion
of a particular ring with the  other rings
not excited.
   This is {\it not}
a self-consistent limit because 
gravitational interactions will
in general excite all of the rings.  
  However this limit 
is informative. 
   With ${\cal E}_j \propto \Delta_j
\propto \exp(-i\omega t)$, $j=1,..,N$, where $\omega$
the ring precession frequency, 
we get
%%%%%%%%
$$
\left[(\omega-\Omega)^2 +
E\right]{\cal E} 
+\left[2\Omega (\omega -\Omega )+d~\right]\Delta=0~,\quad~~
$$
\begin{equation}
\left[2\Omega(\omega-\Omega)+d~ \right]{\cal E}
+\left[(\omega-\Omega)^2+ 
D\right]\Delta 
=0~,
\end{equation}
%%%%%%%%
where the $j$ subscripts
are implicit, 
$E \equiv \Omega_{{\cal E}j}^2 -\Omega_j^2-D_{jj}/M_j$,
$D \equiv \Omega_{\Delta j}^2-\Omega_j^2-D_{jj}/M_j$,
and $d \equiv D_{jj}/M_j$. 
   With $w \equiv \omega-\Omega$, we get
%%%%%%%
\begin{equation}
w^4+(D+E-4\Omega^2)w^2-4d\Omega w +ED-d^2=0~.
\end{equation}
%%%%%%%
For the limit of many rings, the terms
involving $d$ become negligible, and (60)
can be readily solved to give four
real roots if $0<ED<(4\Omega^2-D-E)^2/4$.

  Figure 3a shows the radial dependence of
the four  ring
precession frequencies $\omega_\alpha$
($\alpha =1,..,4$) corresponding to the
four modes of oscillation.  These modes
are the analogues of the normal modes
of vibration of a non-rotating 
system (see Lovelace
1998).

  Three of the modes in Figure 3a 
have positive frequencies ($\omega_\alpha >0$
for $\alpha =2,3,4$) 
so that they have forward 
precession with
\begin{equation}
 \varphi_\alpha
=\omega_\alpha t+ {\rm const}~,\quad 
{\rm and}\quad \psi_\alpha =
\omega_\alpha t+ {\rm const}'~.
\end{equation}
These relations follow from (36) and (37) 
because $|{\cal E}|=\epsilon$ 
and $|\Delta|=\delta$ are constants.
  The other mode ($\alpha=1$) has $\omega_1 <0$
and therefore backward precession.
 
    Figure 3b shows that the two modes 
$\alpha =2,3$ have
${\cal E}/\Delta $ small compared with
unity  near 
the inner radius of the disk, $r_1$.
    Note that ${\cal E}/\Delta=0$ corresponds
to a pure azimuthal shift of the ring matter
without a shift of the ring center.
  The mode $\alpha=1$ with
backward precession has ${\cal E}/\Delta \sim 1$.
   As mentioned, ${\cal E}/\Delta =1$ 
corresponds to a rigid shift of the ring center.

   Evaluation of the ring energy for the
four modes using equation (54)
 shows that the modes $\alpha=1,2$ have
{\it negative energy}  whereas
the modes $\alpha=3,4$ have positive energy.
  The negative energy modes are unstable
in the presence of dissipation, for example,
the force due to dynamical friction
(see for example Lovelace 1998).

  Note that for vanishing ring mass
($D \propto d \propto M_j\rightarrow 0$), 
the middle two
roots approach $\omega 
=\Omega \pm [D E (4\Omega^2-E)]^{1/2}$.
Thus for $M_j\rightarrow 0$,
there are only three different roots
of equation (60), $\omega=\Omega$
and $\omega=\Omega \pm (4\Omega^2-E)^{1/2}$.

  Outside of the central region of
a galaxy we have 
$\Omega \sim 1/r$.   
   Consequently,  
the radial dependences of the mode
frequencies
$\omega_\alpha(r) \propto \pm 1/r$  will tend to
``wrap up'' an initially coherent asymmetry
into a tightly wrapped spiral (in
the absence of ring interactions).
  The forward precessing modes ($\alpha=2-4$)
will give a {\it trailing} spiral wave,
$\varphi_\alpha \propto 1/r$, 
(with respect to the azimuthal motion $v_\phi>0$),
whereas the backward precessing mode ($\alpha =1$)
will give a {\it leading} spiral wave,
$\varphi_1 \propto -1/r$.
  The case of mode $\alpha=1$ for
non-interacting rings
was discussed earlier by Baldwin {\it et al.} (1980).

\section{Eccentric Motion of a ``Disk'' of Two Rings}

   Here, we consider the eccentric motion
of a ``disk'' consisting
of two interacting rings, 
 one of mass 
$M_1$ and radius $r_1$, 
and the other of mass $M_2$,
radius $r_2$. 
 The values of $\Omega_j^2$, 
$\Omega_{{\cal E}j}^2$,
$\Omega_{\Delta j}^2$ ($j=1,2$) 
are given by equations
(A11), (43), and (44) with the bulge and halo 
potentials as given in \S 2.1.  
  Thus the present treatment
is {\it self-consistent} (in contrast with the
previous subsection).
 With ${\cal E}_j$ and
$\Delta_j$ proportional to
$\exp(-i\omega t)$, equations (45) and (46) give
$$
\big[-(\omega-\Omega_1)^2+\Omega_1^2-
\Omega_{{\cal E}1}^2 +D_{11}\big]{\cal E}_1 
\quad \quad \quad \quad \quad \quad
$$
\vspace{-0.2in}
$$
\quad \quad \quad \quad 
+~\big[2\Omega_1(\Omega_1-\omega)-D_{11}\big]\Delta_1
$$
\begin{equation}
=
\big[C_{12}({\cal E}_{2}-{\cal E}_1)
 + D_{12} (\Delta_2 -{\cal E}_1)\big]/M_1~,
\end{equation}
% eqn 61
\vspace{0.01in}
$$
\big[-(\omega-\Omega_1)^2+
\Omega_1^2-\Omega_{\Delta 1}^2 +D^\prime_{11}\big]\Delta_1
\quad \quad \quad \quad \quad \quad
$$
$$ \quad \quad \quad \quad
+~[2\Omega_1(\Omega_1-\omega)-D^\prime_{11}\big]{\cal E}_1 
$$
\begin{equation}
=\big[E_{12}(\Delta_2-\Delta_1)
 + D_{12}^\prime ({\cal E}_2-\Delta_1)\big]/M_1~,
\end{equation}
% eqn 62
\vspace{0.01in}
$$
\big[-(\omega-\Omega_2)^2+\Omega_2^2 -
\Omega_{{\cal E}2}^2+D_{22}\big]
{\cal E}_2 \quad \quad \quad \quad \quad \quad
$$
$$
\quad \quad \quad \quad
+~\big [2\Omega_2(\Omega_2-\omega)-D_{22}\big]\Delta_2
$$
\begin{equation}
=\big[C_{12}({\cal E}_{1}-{\cal E}_2)
 + D_{12}^\prime( \Delta_1 - \Delta_2)\big]/M_2~,
\end{equation}
% eqn 63
\vspace{0.01in}
$$
\big[-(\omega-\Omega_2)^2+
\Omega_2^2-\Omega_{\Delta 2}^2+D^\prime_{22}\big]\Delta_2
\quad \quad \quad \quad \quad \quad
$$
$$
\quad \quad \quad \quad 
+~\big[2\Omega_2(\Omega_2-\omega){\cal E}_2-D^\prime_{22}\big] 
$$
\begin{equation}
=\big[E_{12}(\Delta_1-\Delta_2)
 + D_{12}( {\cal E}_1-\Delta_2)\big]/M_2~.
\end{equation}
% eqn 64
For a non-zero solution, the determinant
of the $4\times4$ matrix multiplying
$({\cal E}_1,~\Delta_1,~{\cal E}_2,~\Delta_2)$
must be zero.  
 This leads to an eighth
order polynomial in $\omega$ which
can be readily solved (with Maple R. 5) for the
frequencies of the $8$ modes ($\alpha=1..8$).

%%%%%%%%%%%%%%%%%%%%%%%%%%%%%%%%%%%%%%%%%%
\begin{figure*}[t]
\epsfscale=400
\plotone{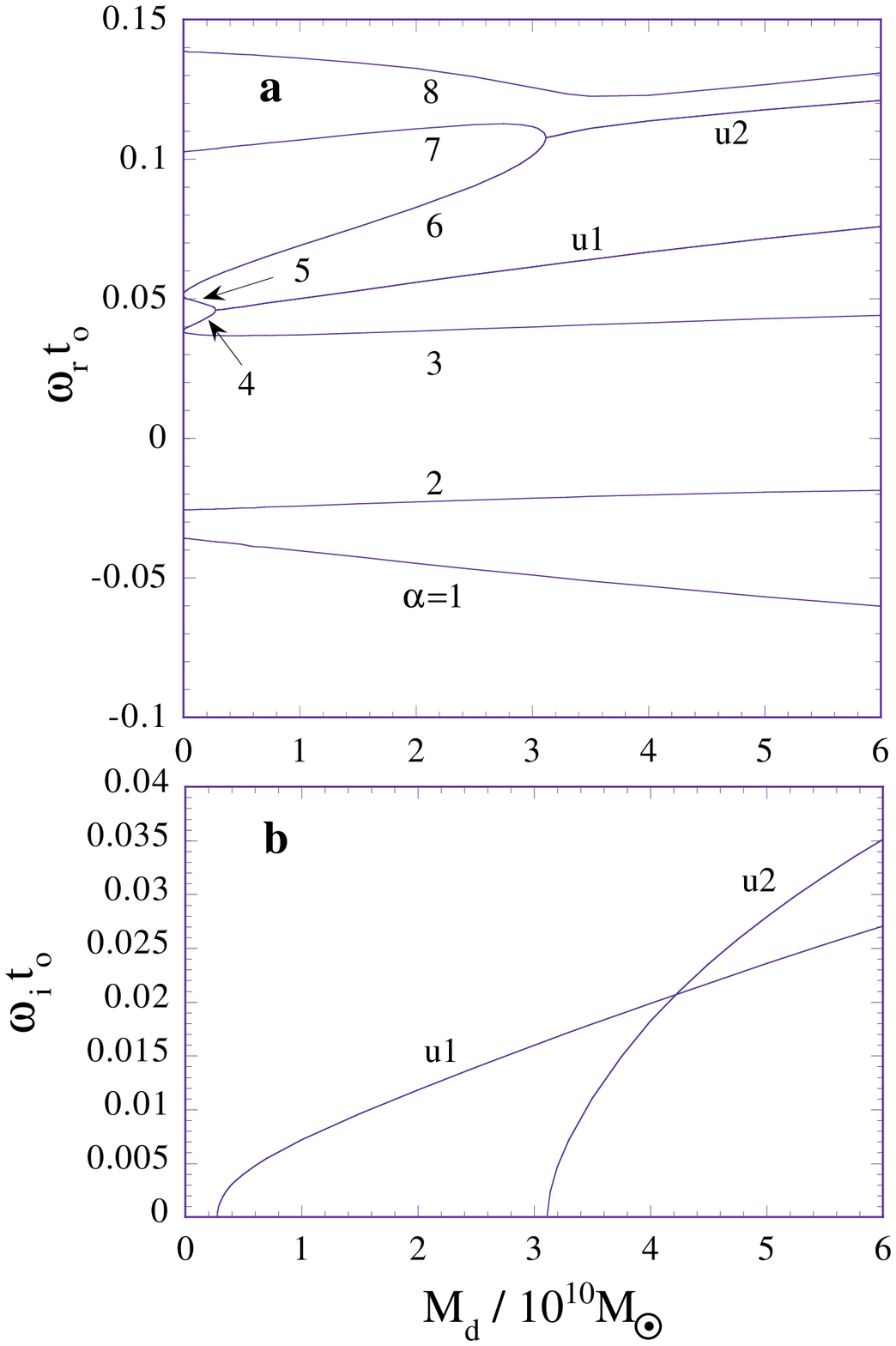}
\caption{
The figure shows
the eccentric instability of a
``disk''  of mass $M_d$ consisting of
two rings of
equal mass $M_d/2$
at equilibrium radii $r_1=3$ 
and $r_2=5$ kpc. 
  The top panel (${\bf a}$)
shows the dependence of the real
parts of the 
frequencies $\omega_{r\alpha}$
($\alpha=1,..,8$) 
on  $M_d$.
  The labels $u1,u2$ indicate
unstable branches where the 
frequency is complex;  $u1$ is
referred to in the text 
as the ``first instability''
and $u2$ the ``second instability.''
   Here, 
$t_o\equiv 10^6$ yr.
   The bulge has
$M_b=0.5\times 10^{10}M_\odot$ and  $r_b=1$ kpc, 
and the halo $v_h=250$ km/s and
$r_h = 5$ kpc in the
expressions given
in \S 2.1.   
  The bottom panel (${\bf b}$) shows
the dependence of the growth rate
$\omega_i$ on the ring mass.  The 
onset of instability corresponds to
the merging of the two real frequencies
in panel (${\bf a}$).  
 The tidal
coefficients, $\{C_{jk}\},$, etc.,  are
obtained using the equations of
the Appendix and 
$\Delta r_j=2$ kpc.
}
\end{figure*}
%%%%%%%%%%%%%%%%%%%%%%%%%%%%%%%%%%%%%%%%%%

  Figure 4 shows the behavior, including
instability, of a system of two rings of
equal mass $M_1=M_2$ so that
the ``disk'' mass is $M_d = 2 M_1$.
   For $M_d \rightarrow 0$, the modes
$\alpha=3,4$ approach $\Omega_2$ and
the modes $\alpha=5,6$ approach $\Omega_1$
which agrees with the behavior found in \S 3.
  Note that for small $M_d$  modes $\alpha=1-4$ are
associated with ring $2$ while modes $\alpha=5-8$
are associated with ring $1$. 
  In the absence of interactions, the
rings are stable independent of their mass. 
   As the disk mass increases, 
the modes $\alpha=4,5$ approach each other
and merge at $M_d\approx 0.276 \times 10^{10}
M\odot$ and give instability with 
${\cal I}m(\omega)\equiv \omega_i >0$ 
as shown in Figure 4b.  
  We refer to this as the ``first instability.''
  Notice that the onset of instability
corresponds to a merging of the positive energy
mode $\alpha=4$ of ring $2$ with the negative
energy mode $\alpha=5$ of ring $1$ (see \S 3).
  The interaction of positive and negative
energy modes is a well-known instability
mechanism (see for example Lovelace, Jore,
\& Haynes 1997).  At the instability
threshold, a dimensionless measure of the
ring self-gravity is 
$GM_d/(\bar{r}^3 \bar{\Omega}^2) \sim 0.1$,
where $\bar{r}=4$ kpc and $\bar{\Omega}
\approx 0.0458/t_o$.
    The dependence
of the growth rate is well fitted
by $\omega_i t_o 
\approx 0.00842(M_d-M_{c1})^{0.67}$,
with the masses in units of $10^{10}M_\odot$
and $t_o \equiv 10^6$ yr.
   For $M_d >3.12\times 10^{10}\equiv M_{c2}$,
there is a ``second instability'' with growth
rate $\omega_it_o \approx 
0.0196(M_d-M_{c2})^{0.55}$.

   The ratios of the complex perturbation 
amplitudes can readily be obtained from
(62) - (65) once the $8$, possibly
complex frequencies are known.
  For the ``first instability,'' for example,
for $M_d=  10^{10}M_\odot$ and the same
conditions as for Figure 4,
we find ${\cal E}_1 \approx 0.254-0.117i$,
$\Delta_1=1$ (by choice), ${\cal E}_2 \approx
0.0108-0.365i$, and $\Delta_2 \approx
0.673+1.30i$.  
  These values correspond to
$\varphi_1 \approx 24.7^\circ$, 
$\varphi_2 \approx 91.7^\circ$,
$\psi_1=0$ (by choice), and
$\psi_2\approx 297^\circ$.
  Thus the azimuthal density enhancement
in the outer ring {\it trails} the
density enhancement of the inner ring. 
  On the other hand,  
 the radial shift of the outer ring
{\it leads} the shift of the inner ring.

  As a second example, for 
$M_d =4 \times 10^{10}M_\odot$ 
both the ``first'' and ``second'' 
instability occur.  
  For the
``first'' instability, we again find that
the azimuthal density enhancement
of the outer ring {\it trails} that
of the inner ring, whereas 
the radial shift of the 
outer ring {\it leads} that
of the inner ring. 
   For the ``second'' instability
the situation is different in
that {\it both} the azimuthal density
enhancement and the radial shift
of the outer ring {\it trail} those
in the inner ring.  
  Specifically,
we find $\varphi_1\approx 178^\circ$,
$\varphi_2 \approx 89.3^\circ$, 
$\psi_1=0$ (by choice), 
and $\psi_2 \approx 283^\circ$.
  Note that for both rings, the angle of 
the radial shift is roughly $180^\circ$
displaced from the
azimuthal density enhancement.  
  The displacement of the center of mass
of the ring (equation 20) is 
dominated by the azimuthal
displacement ($|{\cal E}_1|
\approx 0.70|\Delta_1|$
and $|{\cal E}_2|\approx 0.67|\Delta_2| 
\approx 0.22|\Delta_1|$).
   Thus having $\varphi_j$
and $\psi_j$ about $180^\circ$ out of phase  
allows the center of mass of
each ring to be closer to the origin,
and this is a lower energy configuration.

 Consider now the angular momentum of
the perturbed two ring system 
which from (58) is
 ${\cal P}_\phi = {\cal P}_{1\phi}
+{\cal P}_{2\phi}={\rm const}$.   
   For the evaluation of
${\cal P}_\phi$, note that for each
of the eight modes,
$\dot{\Delta}_j = -i\omega_\alpha \Delta_j$,
and $\dot{\cal E}_j=-i\omega_\alpha {\cal E}_j$,
where $\alpha=1,..,8$ labels the mode.
 This implies  that 
%%%%  eqn 65
\begin{equation}
\omega_\alpha
=\dot{\psi}_j +i(\dot{\delta}_j/\delta_j)
~=~\dot{\varphi}_j +i(\dot{\epsilon}_j/\epsilon_j)~.
\end{equation}
 Thus, for the ``first instability,'' 
for $M_d= 10^{10}M_\odot$ where $\omega t_o
\approx 0.0500+0.00719i$,
we have $\dot{\psi}_j=\dot{\varphi}_j=0.05/t_o$,
(which corresponds to forward precession), and
$\dot{\delta}_j/\delta_j =\dot{\epsilon}_j
/\epsilon_j = 0.00719/t_o$.
  For an unstable mode, the coefficients
of the six terms in ${\cal P}_\phi$ are all 
grow exponentially.  
   The only possible
way in which ${\cal P}_\phi =$const can
be maintained is to have ${\cal P}_\phi =0=
{\cal P}_{\phi1}+{\cal P}_{\phi2}$.
   This relation provides a useful check
on the correctness of the calculations.
    
  For the ``first instability'' ($M_d>M_{c1}$),
we find by evaluating
(58) that 
${\cal P}_{\phi1}=
-{\cal P}_{\phi2} >0$.
   This means that
the angular momentum of the inner ring increases
while that of the outer ring decreases.
  Thus the  instability
acts to transfer angular momentum {\it inward}.
  In contrast, for the ``second instability''
($M_d>M_{c2}$), we find that
${\cal P}_{\phi1}=-{\cal P}_{\phi2} < 0$.
  Thus, the ``second instability'' 
acts to transfer angular momentum
{\it outward}.  
  If $P_{\phi1}$ decreases and $P_{\phi2}$
increases, then the average radius of ring
1 must decrease and 
that of ring 2 must increase.
   Therefore, the ``second instability'' may be
important for the accretion of matter in
a gravitating disk.

\section{Eccentric Motion of 
One Ring and ``Center''}

    Here, we consider the eccentric motion of
a ``disk'' of one ring including the influence
of the eccentric motion of the ``center'' $M_0$
which is located at ${\bf r}=0$ in equilibrium.
  The mass $M_0$ includes the mass of a central
black hole $M_{bh}$ if it is present.
   The ring perturbation is described by ${\cal E}(t)$
and $\Delta(t)$ as given by (52) and (53), while
the ``center'' is described by ${\cal E}_0(t)$ which
is given by (49).   
  With the perturbations
$\propto \exp(-i\omega t)$, we find
%%%%%%%%
$$
\left[(\omega-\Omega)^2 +
E\right]{\cal E} 
+\left[2\Omega (\omega -\Omega )+d~\right]\Delta
=2\Omega_a^2{\cal E}_0~,
$$
$$
\left[2\Omega(\omega-\Omega)+d~ \right]{\cal E}
+\left[(\omega-\Omega)^2+ 
D\right]\Delta 
=~\Omega_a^2 {\cal E}_0~,
$$
\begin{equation}
(\omega^2-\Omega_0^2){\cal E}_0=
\Omega_b^2 \bigg({\cal E}-{\Delta \over 2}\bigg)~,
\end{equation}
%%%%%%%%
where $\Omega_a^2 \equiv G M_0/r_1^3$ and
$\Omega_b^2 \equiv G M_1/r_1^3$, with
$M_1$ and $r_1$  the mass and radius of the
ring.  
  For a non-zero solution, the determinant
of the $3\times 3$ matrix multiplying
$({\cal E}, \Delta, {\cal E}_0)$ must
be zero.   
  This leads to a sixth order
polynomial in $\omega$ or $w \equiv \omega-\Omega$
which can readily be solved (with Maple, R5)
for the frequencies of the $6$ modes
$\omega_\alpha$ ($\alpha=1..6$).
   We obtain 
$$
\left [\left (w+\Omega\right )^{2}
-{{\Omega_0}}^{2}\right]\left [\left ({w}^{2}+
E\right )\left ({w}^{2}+D\right )-\left (2
\Omega w+d\right )^{2}\right]
$$
\begin{equation}
-\Omega_{ab}^4\left ({\frac {
5{w}^{2}}{2}}+2D+{\frac {E}{2}}+2\Omega w
+d\right )=0~,
\end{equation}
where the strength of the interaction
between the ring and the ``center'' is
measured by 
% eqn 69
\begin{equation}
\Omega_{ab}^2 \equiv
\Omega_a \Omega_b ={ G \sqrt{M_0 M_1}\over r_1^3}~.
\end{equation}
Here, $D,~E,~d$ and $\Omega=\Omega_1$ are defined in 
(59), and $\Omega_0$ is defined in (51).

   Figure 5 shows the dependence of 
the growth rate 
$\omega_i={\cal I}m(\omega_\alpha)$
on the mass of the ``center'' $M_0$ with the
mass of the ring held fixed.  
  The associated real part of the
frequency is positive.
  The onset of instability  corresponds
to the point where two of the six modes
with frequencies given by (68) merge. 
  The merging is again of positive and
negative energy modes.
  The growth rate has to
a good approximation
the dependence
$\omega_i t_o \approx 0.00764 (M_0-2.61)^{1/2}$,
with $M_0$ in units of $10^9 M_\odot$
and $t_o = 10^6$ yr. 

%%%%%%%%%%%%%%%%%%%%%%%%%%%%%%%%%%%%%%%%%%
\begin{figure*}[t]
\epsfscale=500
\plotone{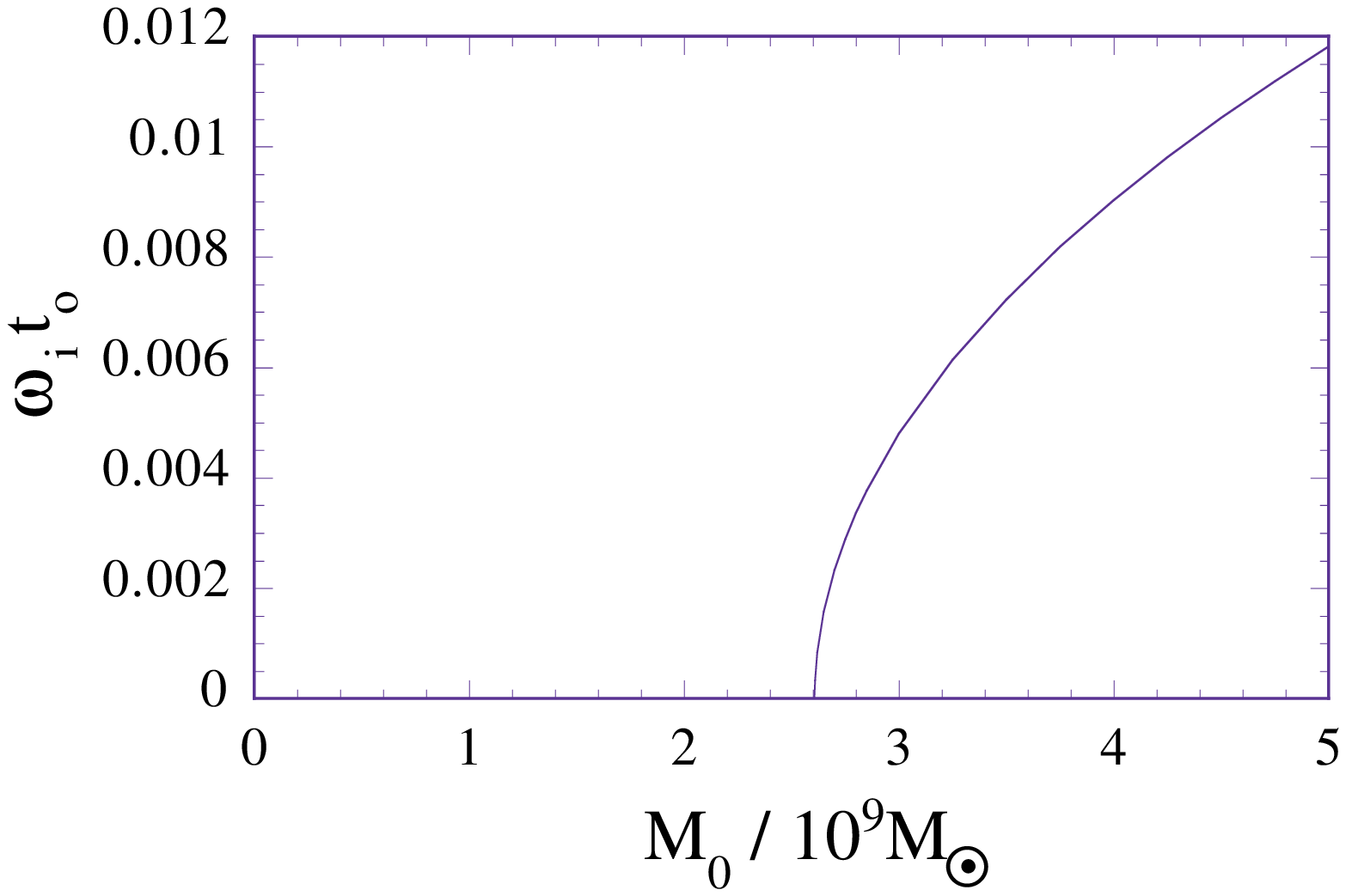}
\caption{
The figure shows
the eccentric instability of a
ring of mass $M_1$  
interacting gravitationally 
with a displaced point
mass $M_0$ shifted from its
equilibrium position ${\bf r}=0$. 
   The halo and bulge potentials
are the same as in Figure 1.
  The mass $M_0$ includes
the mass of a central black
hole $M_{bh}$ if present.
  The mass of the ring is 
$M_1= 6 \times 10^{10} M_\odot$
and its radius is $r_1 =3$ kpc.
   Here, 
$t_o\equiv 10^6$ yr.  The tidal
coefficients are evaluated
using the expressions of the
Appendix with $\Delta r_1=2$ kpc.
}
\end{figure*}
%%%%%%%%%%%%%%%%%%%%%%%%%%%%%%%%%%%%%%%%%%

  The ratios of the complex amplitudes
follow from (67), and for the unstable
mode for $M_0 =2.75 \times 10^9M_\odot$
we find $\omega t_o \approx 0.155 +0.00289i$,
$\Omega t_o \approx 0.131$, $\Omega_0 t_o 
\approx 0.142$, $\Omega_{ab} \approx 0.0465$, 
${\cal E} \approx -0.817+0.0333i$, 
$\Delta =1$ (by choice), and ${\cal E}_0
\approx -3.38+0.916i$ which give
$|{\cal E}|\approx 0.817$, $|{\cal E}_1| 
\approx 3.51$, $\varphi \approx 182^\circ$,
and $\varphi_0 \approx 195^\circ$, where
$t_o=10^6$ yr. 
    Thus, the radial shift of the ring is roughly
$180^\circ$ away from the maximum of the
density enhancement which 
has $\psi=0$ (by choice).
  The shift of the center of mass of the
ring (equation 20) is dominated by the
azimuthal density enhancement. 
   Thus the radial shift and azimuthal
displacements are such that the center
of mass of the ring moves closer to
the origin which is a lower energy
configuration.
   Note that the radial shift of 
the ``center'' {\it trails}
the  ring center of mass
by an angle $360^\circ-\varphi_0 \approx 165^\circ$.
  Hence, the torque of the ``center'' on the ring
acts to {\it reduce} the angular momentum of the
ring  as verified below.  At the same time
the ring acts to increase the angular momentum
of the ``center.''

  Consider now the angular momentum of the
perturbed ring plus ``center'' 
system which is given
by (58).   
  Following the
arguments of the prior section,
the sum of the angular momentum of
the ``center'' and that of the
ring must be zero for a growing
mode.  
   The angular momentum of the
``center'' is simply 
$M_0\epsilon_0^2 \dot{\varphi}_0$.
  For a pure mode, we have 
${\cal R}e(\omega_\alpha)=\dot{\varphi}_0$.
  As mentioned, the real part of the
frequency is positive for the
unstable mode and therefore the angular
momentum of the ``center'' increases
while the angular
momentum of the ring decreases.
  Thus, there is a transfer of angular
momentum from the ring to the ``center.''
  Due to the loss of angular momentum 
the average radius of the ring will
decrease.  Thus the instability may
be important for accretion of matter
to the galaxy center.

\section{Eccentric Motion of N Rings}

   For the results presented here,
the rings are taken to 
be uniformly spaced in $r$
with radii $r_j = 1 +(j-1)\delta r$ kpc  
with $\delta r =0.5$ kpc
and with $j=1,..,N=31$. 
  The value $N=31$ gives good
spatial resolution over all but
the inner part of the disk.
    The outer
radius $r_N$ (in the range
say $10 -20$ kpc) has 
little influence on the eccentric
motion described here, as  
   verified  by comparing
results with $r_N=16$ kpc 
those obtained with
significantly larger  $r_N$.
  Also, the eccentric motion of
the outer disk, say, $r \gtrsim 3$ kpc, is
essentially independent of $\delta r$.
   We first consider in \S 6.1 the case
where the ring masses correspond to
the exponential distribution
discussed in \S 2.1. 
   The inner part of the disk is found
to be strongly unstable to eccentric 
motions and
therefore in \S 6.2 we consider a disk
with the mass of the innermost three rings
reduced.   In \S 6.3 we consider disks
with a smooth reduction in $\Sigma_d(r)$ 
in the inner  part of the disk, $r \lesssim r_d$.

  We solve   
(52) and (53) numerically 
as eight first order
equations for 
$\epsilon_{xj},~\dot{\epsilon}_{xj},~\epsilon_{yj},$
and $\dot{\epsilon}_{yj}$, and for 
$\delta_{xj},~\dot{\delta}_{xj},~\delta_{yj},$
and $\dot{\delta}_{yj}$, $j=1,..,N$.  
 At the same time, we solve the two additional
equations,
\begin{equation} 
{d\varphi_j \over dt~}= { \epsilon_{xj}\dot{\epsilon}_{yj} -
\epsilon_{yj}\dot{\epsilon}_{xj}
\over \epsilon_{xj}^2+\epsilon_{yj}^2}~,
\end{equation}
\begin{equation}
{d\psi_j \over dt~}= { \delta_{xj}\dot{\delta}_{yj} -
\delta_{yj}\dot{\delta}_{xj}
\over \delta_{xj}^2+\delta_{yj}^2}~,
\end{equation}
to give $\varphi_j(t)$, which is the
angle to the maximum of the radial shift, and
$\psi_j(t)$, which is the angle to the maximum
of the azimuthal density enhancement.
  These angles are analogous
to the line-of-nodes angles for
the tilting of the rings of a disk
galaxy (Lovelace 1998).
  Thus, we solve $10N$ first order equations.
   In all cases, the
total energy (54)
and total canonical 
angular momentum (58)
are accurately conserved.
  The different frequencies $\Omega_j$,
$\Omega_{{\cal E}j}$, and $\Omega_{\Delta j}$,
and the tidal coefficients 
$\{C_{jk}\}$, etc., are
evaluated using the equations 
of the Appendix.

\subsection{Exponential Disk}

    Here, we consider the eccentric
motion of the rings for the case
where
$M_j=2\pi r_j \delta r
\Sigma_d(r_j)$ with 
$\Sigma_d(r)=\Sigma_{d0}\exp(-r/r_d)$. 
  The mass of the center is
assumed given by (47), which
gives $M_0\approx 1.06 \times 10^9$
for the parameters of Figure 1. 
  Alternatively, this value of $M_0$
could be due in part to a central black
hole.

   Figure 6 shows the dependences
 of the radial shifts $\epsilon_j(\varphi_j)$
 and azimuthal
displacements $\delta_j(\psi_j)$ of
the rings ($j=1-31$) and the radial
shift of the
``center'' $\epsilon_0(\varphi_0)$ at
a short time, $100$ Myr after
an initial perturbation.  
  This type of plot is related to the
plots emphasized by Briggs (1990) 
for characterizing the warps of 
galactic disks (see also Lovelace 1998).  
   The angles 
$\varphi$ and $\psi$ are
analogs of the line-of-nodes
angle the warp.

%%%%%%%%%%%%%%%%%%%%%%%%%%%%%%%%%%%%%%%%%%
\begin{figure*}[t]
\epsfscale=500
\plotone{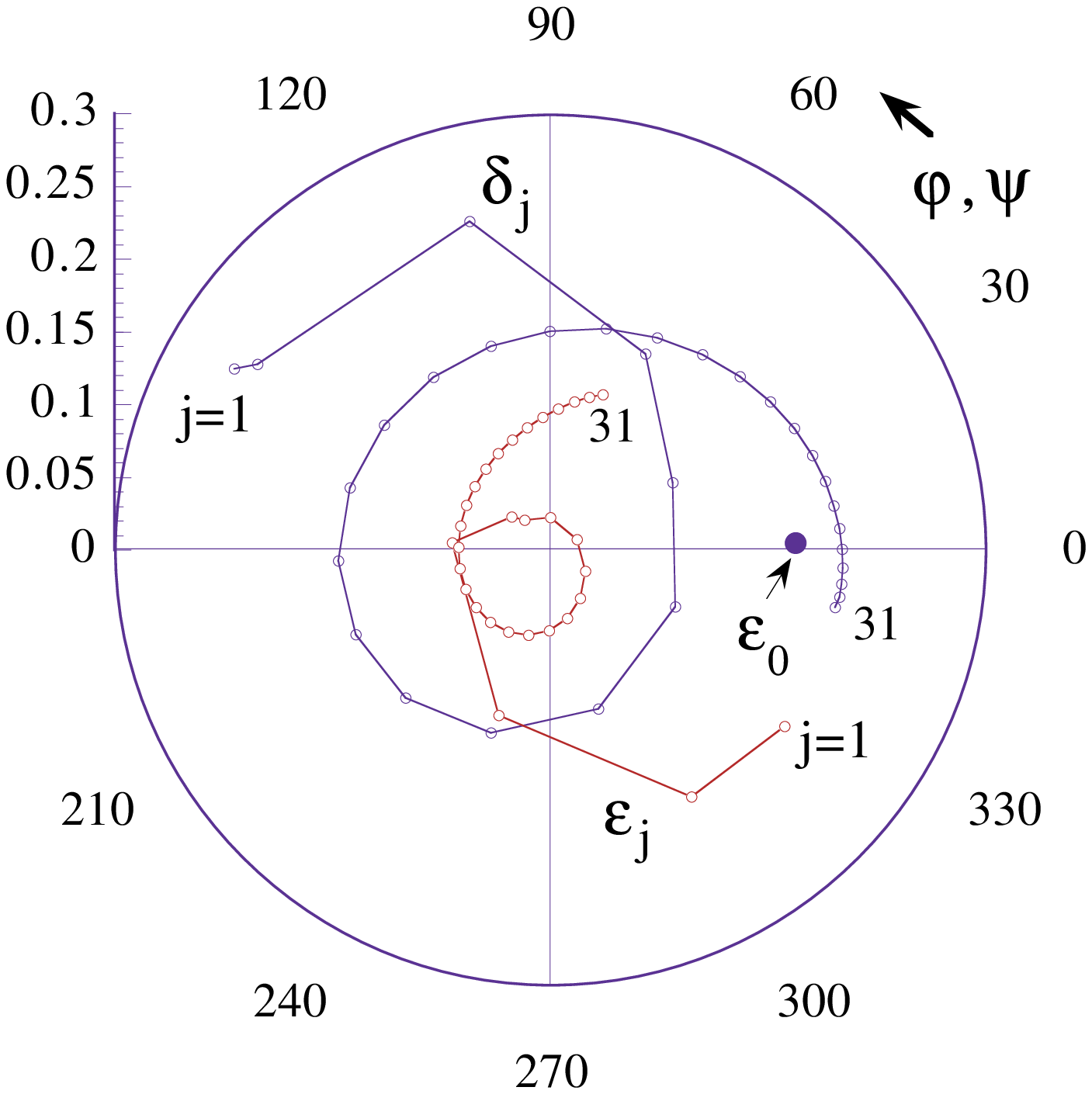}
\caption{
   Polar plot of the radial
shift $\epsilon_j$ and
azimuthal displacement $\delta_j$
of ring matter as a function of 
the angles $\varphi_j$ and $\psi_j$
($j=1-31$)
at time $t=100$ Myr.
   The radial shift of the ``center''
 $\epsilon_0$ is indicated by the
solid dot.
  The  conditions correspond
to the galaxy parameters of Figure 1.
   The rings have radii
$r_j=1+\delta r(j-1)$ kpc and $\delta r=0.5$ kpc 
for $j=1,..,31$,
masses $M_j=2\pi r_j \delta r \Sigma_d(r_j)$
with $\Sigma_d$ given in \S 2.1,
and $M_0$ given by (47), 
which gives $M_0 \approx 1.06\times 10^9 M_\odot$.
  The initial values of the shifts
and displacements are 
$\epsilon_j =0.1(r_j/r_{max})=\delta_j$,
$\varphi_j=0=\psi_j$, and 
$\epsilon_{0x} =
10^{-6}$ and $\epsilon_{0y}=0$. 
  The units of $\epsilon_j$
and $\delta_j$ are arbitrary in that
the equations are linear.
}
\end{figure*}
%%%%%%%%%%%%%%%%%%%%%%%%%%%%%%%%%%%%%%%%%%

%%%%%%%%%%%%%%%%%%%%%%%%%%%%%%%%%%%%%%%%%%
\begin{figure*}[b]
\epsfscale=500
\plotone{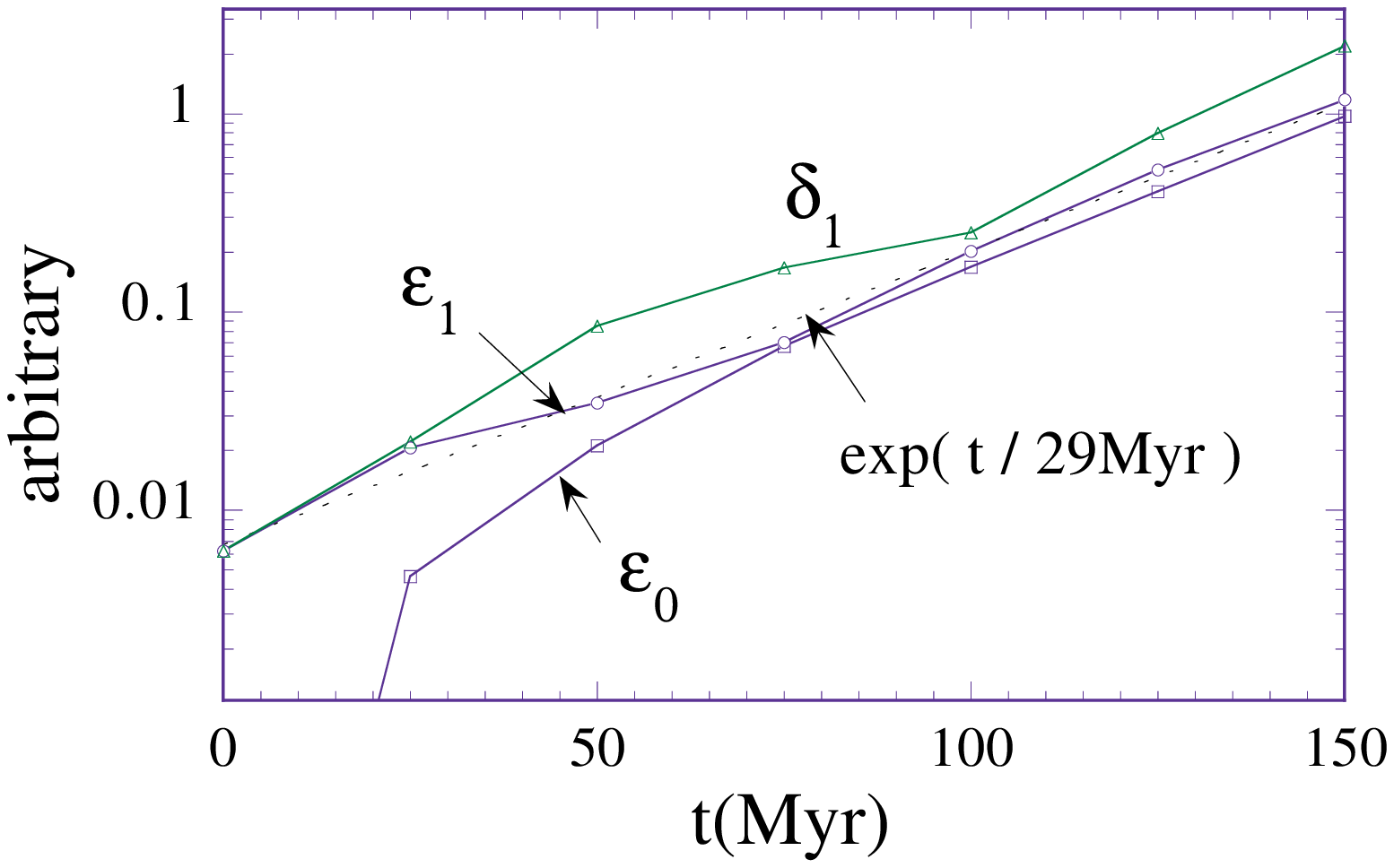}
\caption{
The plot shows
the exponential growth of center
shift $\epsilon_0$ and
radial shift $\epsilon_1$ and
azimuthal displacement $\delta_1$ of
the first ring at $r_1=1$ kpc for
the same conditions as for
Figure 6. 
}
\end{figure*}
%%%%%%%%%%%%%%%%%%%%%%%%%%%%%%%%%%%%%%%%%%

  From Figure 6 note
that the azimuthal displacements
$\delta_j$ are larger than
the radial displacements 
$\epsilon_j$ so that the
displacement of the center 
of mass of a ring is 
dominated by $\delta_j$. 
   The eccentric motion of the
inner rings, say, $j=1-4$ or $r_1=1$ to
$r_4 =2.5$ kpc,
show the most rapid, exponential
growth.   
   For these rings
the angles $\psi_j$ and $\varphi_j$
are approximately $180^\circ$
degrees out of phase, and
this agrees with the behavior
found for the ``second'' instability
of two rings discussed in \S 4.
  As mentioned, this allows the
center of mass of each ring
to move closer to the origin,
which is a lower energy
configuration.
  Note that with increasing
$j$, both $\varphi_j$ and
$\psi_j$ decrease for the
inner rings which corresponds
to a {\it trailing} pattern
the same as found for the
second instability of two rings.
  Note also that the shift
of the ``center'' $\epsilon_0$
trails the shift of the center
of mass of the $j=1$ ring
in agreement with \S 5.
  Thus the torque of the first 
ring on the center acts to
increase the angular momentum
of the center while the
torque of the center on the
first ring decreases the 
rings angular momentum.

    Figure 7 shows the 
exponential growth of the
azimuthal displacement $\delta_1$ and
radial shift $\epsilon_1$ of the first
ring and the simultaneous
growth of the radial shift
of the center $\epsilon_0$.
  The $e-$folding time is about $29$ Myr.
  For comparison, the period of
oscillation of the center is
$T_0 =2\pi/\Omega_0 \approx 46$
Myr for the conditions shown,
where $\Omega_0$ is given
by (51).
   The growth of the eccentric
motion of the inner rings is
reduced somewhat if the mass
of the center is reduced 
to $M_0 = 10^8 M_\odot$;
the $e-$folding time for in this
case is about $38$ Myr for ring 1.
   Figure 8 shows the 
perturbations of
the angular momentum of the center
$P_0$
and the rings $P_j$ 
at $t=100$ Myr 
obtained from (58). 
   In agreement with \S 5
and the abovementioned direction
of the torque,
the angular momentum of the
center increases while that
of the first and second ring
decrease.  
    The decrease in angular
momenta of these rings will
result in their radii shrinking. 
   Note that the center rotates
in the same direction as the
disk matter.

%%%%%%%%%%%%%%%%%%%%%%%%%%%%%%%%%%%%%%%%%%
\begin{figure*}[t]
\epsfscale=500
\plotone{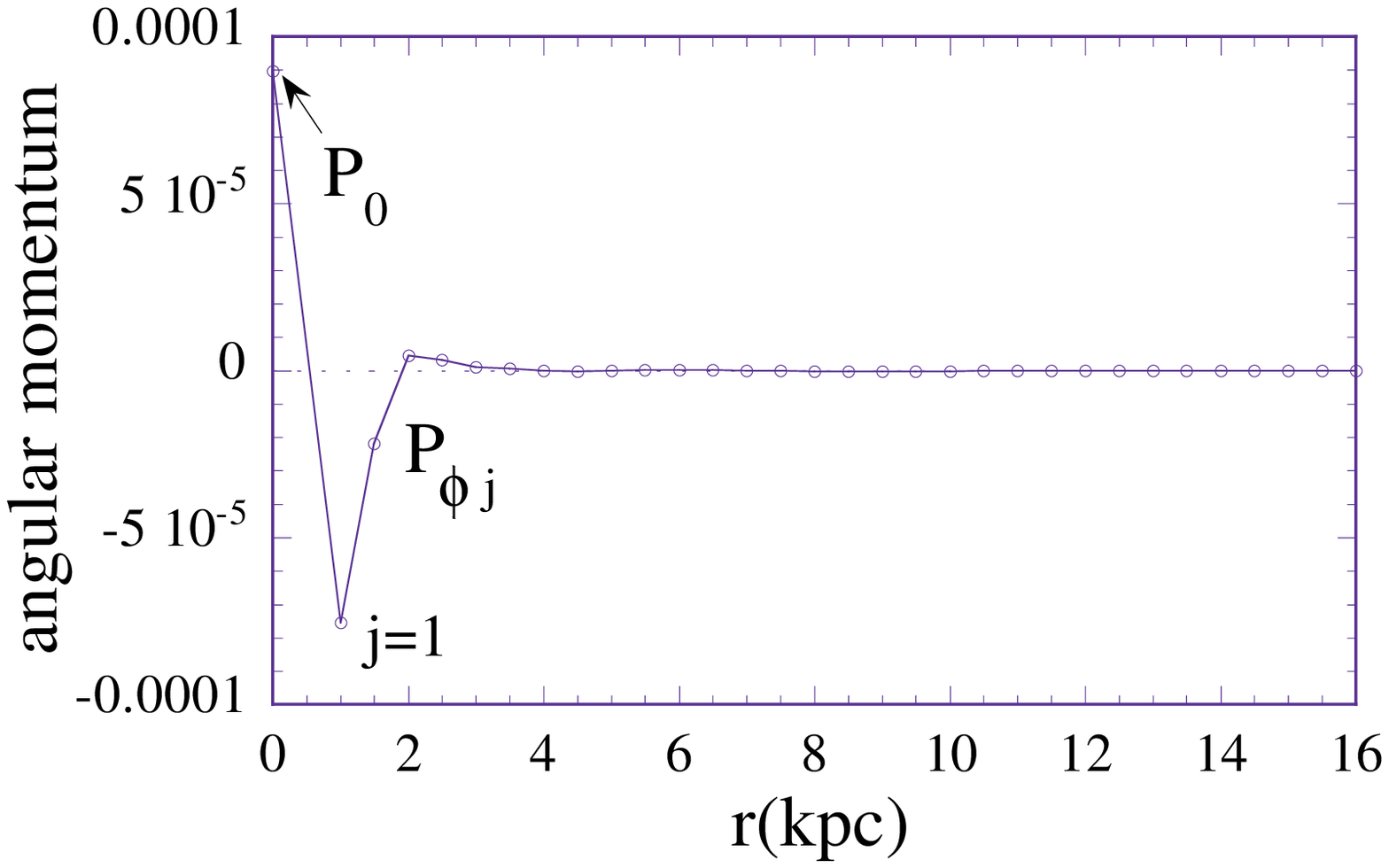}
\caption{
  Plot of
the perturbations of the
angular momentum of the center
$P_0$
and the rings $P_j$ 
at $t=100$ Myr
for the same conditions as Figure 6.
}
\end{figure*}
%%%%%%%%%%%%%%%%%%%%%%%%%%%%%%%%%%%%%%%%%%

   The present
linear theory does not address
the issue of saturation of 
growth of the eccentric motion.
   One possibility is that 
the strong  instability
of the inner rings of the disk
leads to the destruction of this
part of the disk.

\subsection{Exponential Disk with
Rings 1-3 Reduced}

    Here, we consider the eccentric
motion of the rings for the case
where the ring masses 
are the same
as in \S 6.1 {\it except} that
$M_1 \rightarrow 10^{-2} M_1$,
$M_2 \rightarrow 0.1 M_2$, and
$M_3 \rightarrow 0.3 M_3$,
which are the rings with
radii $r_1=1$, $r_2=1.5$, and
$r_3=2$ kpc.  The mass
of the center is the same as
in \S 6.1, 
$M_0\approx 1.06 \times 10^9 M_\odot$.
   The disk mass is reduced by a factor
$\approx 0.84$ compared with an
exponential disk.
  The aim  of reducing $M_1-M_3$
is to reduce
the growth rate of the eccentric motion
of this part of the disk. 

%%%%%%%%%%%%%%%%%%%%%%%%%%%%%%%%%%%%%%%%%%
\begin{figure*}[b]
\epsfscale=500
\plotone{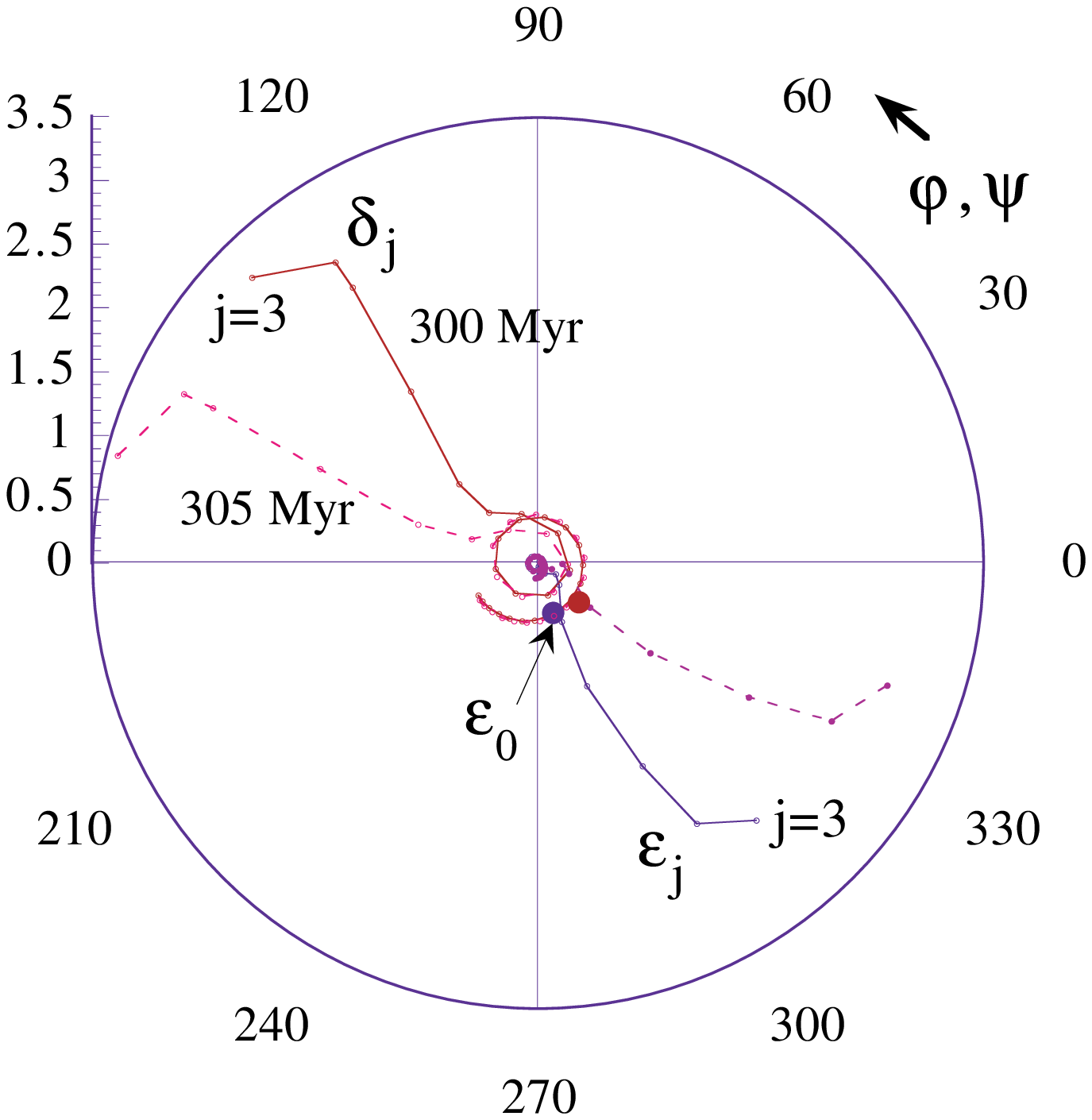}
\caption{
   Polar plot of the radial
shift $\epsilon_j$ and
azimuthal displacement $\delta_j$
of ring matter as a function of 
the angles $\varphi_j$ and $\psi_j$
of the maxima of the shift and
displacement at times $t=300$ and $305$ Myr.
   The solid dot labeled by the
arrow indicates the shift of the
center at $t=300$ Myr;  the other
dot is the shift at $305$ Myr.
   The rings, the center, and
the initial values of the shifts
and displacements are the
same as in Figure 6 except that
$M_1 \rightarrow 10^{-2}M_1$,
$M_2\rightarrow 0.03M_2$, and 
$M_3 \rightarrow 0.1M_3$.  
   Thus the conditions correspond
to the galaxy parameters of Figure 1
except that the disk mass is reduced
to $\approx 5.06 \times 10^{10}M_\odot$.
 The shifts and displacements of the
rings $j=1,2$ are dynamically unimportant
and are not shown.
}
\end{figure*}
%%%%%%%%%%%%%%%%%%%%%%%%%%%%%%%%%%%%%%%%%%

   Figure 9 shows the essential
behavior in a polar plot
of the displacements and shifts
at two times.  
   The radial shift of the
center is negligible on
the scale of this figure.
   The $e-$folding time for
ring 3 is about $49$ Myr.
  Notice that the curves
traced out by $\delta_j$
and by $\epsilon_j$ are
approximated straight
lines from the origin
which rotate
rigidly in 
the direction of motion 
of the matter for
$j=4$ to about $j=11$,
which corresponds to
$r_3=2.5$ to $r_{11}=6$ kpc.
  The instantaneous period 
of rotation of
this pattern is
$\approx 60$ Myr, which is
longer that the 
oscillation period at the
center, $T_0 \approx 46$ Myr.
   This case is an example of the
{\it phase-locking} of the
eccentric motion of these
rings due to the self-gravity
between the rings.  
  This phase-locking is analogous
to that which occurs in
the tilting motion of the
rings representing a disk galaxy due
to self-gravity (Lovelace 1998). 
   In the case of tilting of 
rings the phase-locking results
in the line-of-node angles of
the rings in the inner part of
the disk becoming the same.

   Also in this case the growth
of the eccentric instability
of the inner rings is
sufficiently fast that it probably
further disrupts the inner
part of the disk.

\subsection{Reduced Inner Disk}

    Here, we study the eccentric motion
of a disk the inner part of which
is attenuated relative to an 
exponential disk.
Specifically, the rings masses are
$M_j=2\pi r_j\delta r$ $\hat{\Sigma}_d(r_j)$,
where 
\begin{equation}
\hat{\Sigma}_d(r)=
\Sigma_{d0}\exp\left(-\beta{r_d \over r}\right)
\exp\left(-{r \over r_d}\right)~,
\end{equation}
with $\beta=$const.  
  We consider  $\beta=1$.
  The mass
of the center is the same as
in \S 6.1, 
$M_0\approx 1.06 \times 10^9 M_\odot$.
  However, the motion of
the center is negligible and
 value $M_0$ has 
little influence on the
eccentric motion of the disk described below.
   The disk mass is smaller than
for an  exponential disk
by a factor $\approx 0.47$.

%%%%%%%%%%%%%%%%%%%%%%%%%%%%%%%%%%%%%%%%%%
\begin{figure*}[t]
\epsfscale=500
\plotone{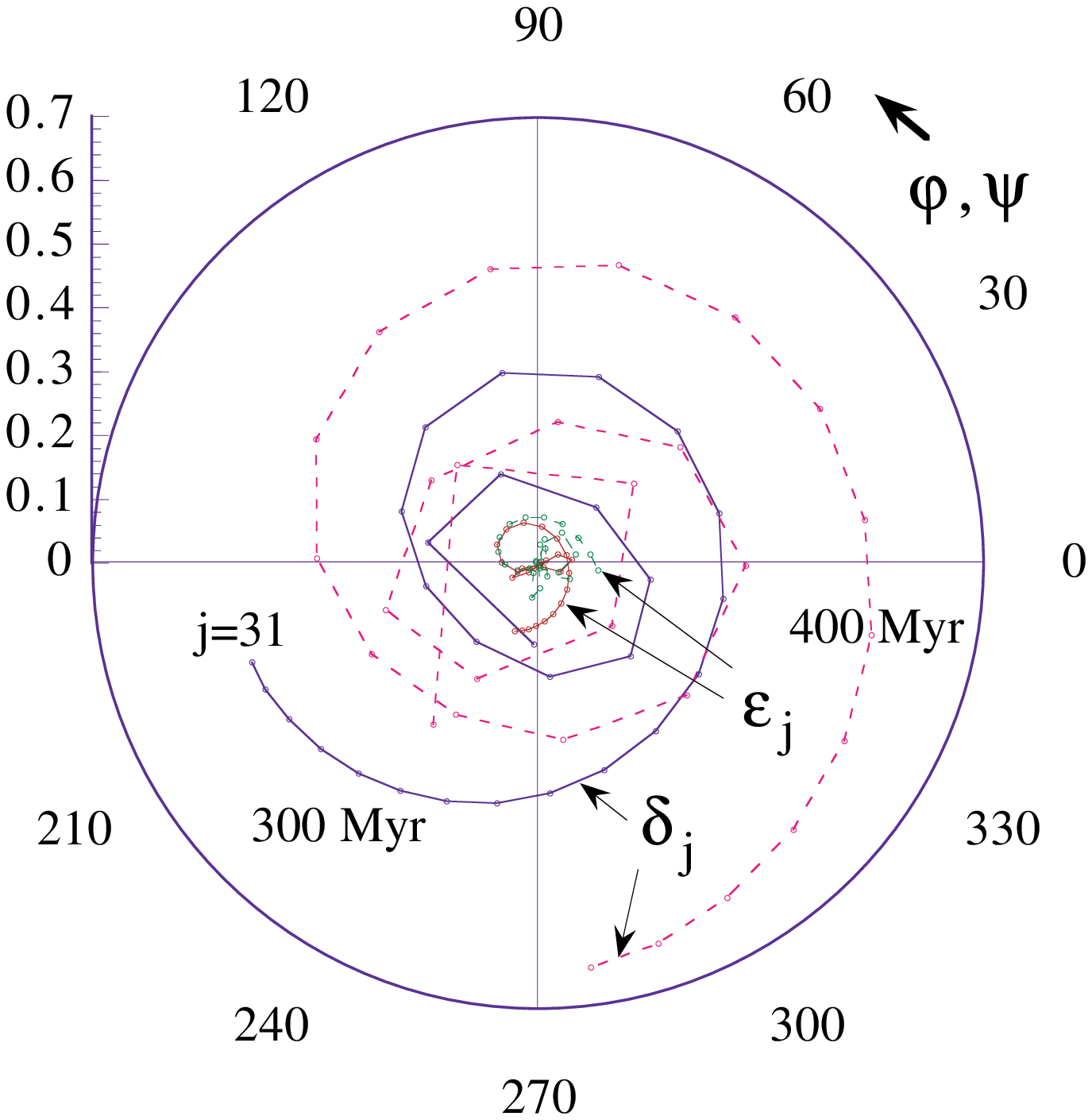}
\caption{
   Polar plot of the radial
shift $\epsilon_j$ and
azimuthal displacement $\delta_j$
of ring matter as a function of 
the angles $\varphi_j$ and $\psi_j$
of the maxima of the shift and
displacement at times $t=300$ and $400$ Myr.
   The radial shift of the center is
negligible on the scale of this plot.
   The rings, the center, and
the initial values of the shifts
and displacements are the
same as in Figure 6 
except that the ring masses
are obtained from (72). 
  The mass of the center is
the same as in Figure 6, $M_0\approx
1.06\times 10^9M_\odot$.
  Thus the conditions correspond
to the galaxy parameters of Figure 1
except that the disk mass is reduced
to $\approx 2.82 \times 10^{10}M_\odot$.
 The shifts and displacements of the
rings $j=1,2$ are dynamically unimportant
and are not shown.
}
\end{figure*}
%%%%%%%%%%%%%%%%%%%%%%%%%%%%%%%%%%%%%%%%%%

%%%%%%%%%%%%%%%%%%%%%%%%%%%%%%%%%%%%%%%%%%
\begin{figure*}[b]
\epsfscale=500
\plotone{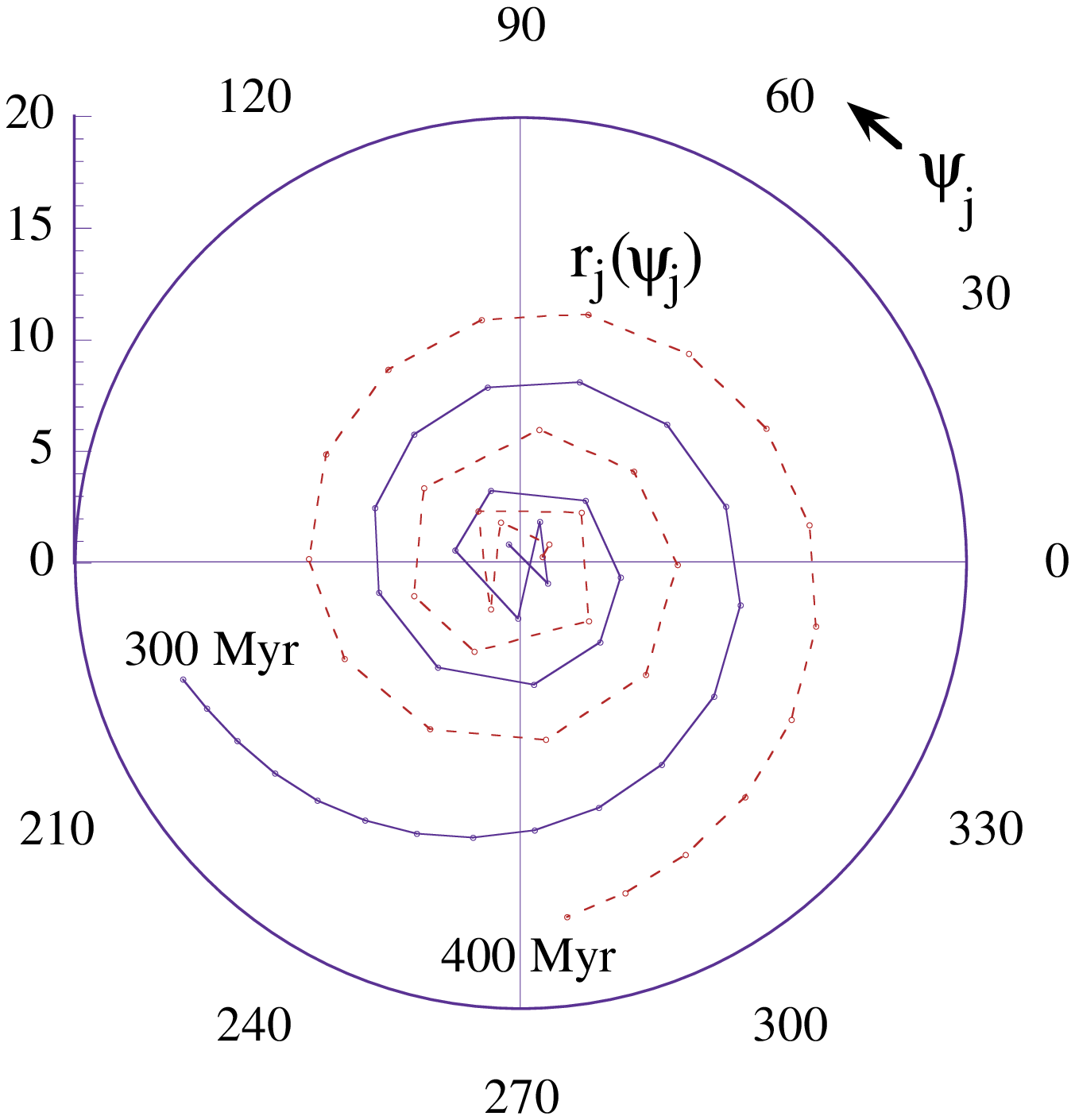}
\caption{
 Polar plot of the radius
to the maximum of the azimuthal
displacement $r_j(\psi_j)$ at
two times 
for the same case as Figure 10.
}
\end{figure*}
%%%%%%%%%%%%%%%%%%%%%%%%%%%%%%%%%%%%%%%%%%

%%%%%%%%%%%%%%%%%%%%%%%%%%%%%%%%%%%%%%%%%%
\begin{figure*}[t]
\epsfscale=500
\plotone{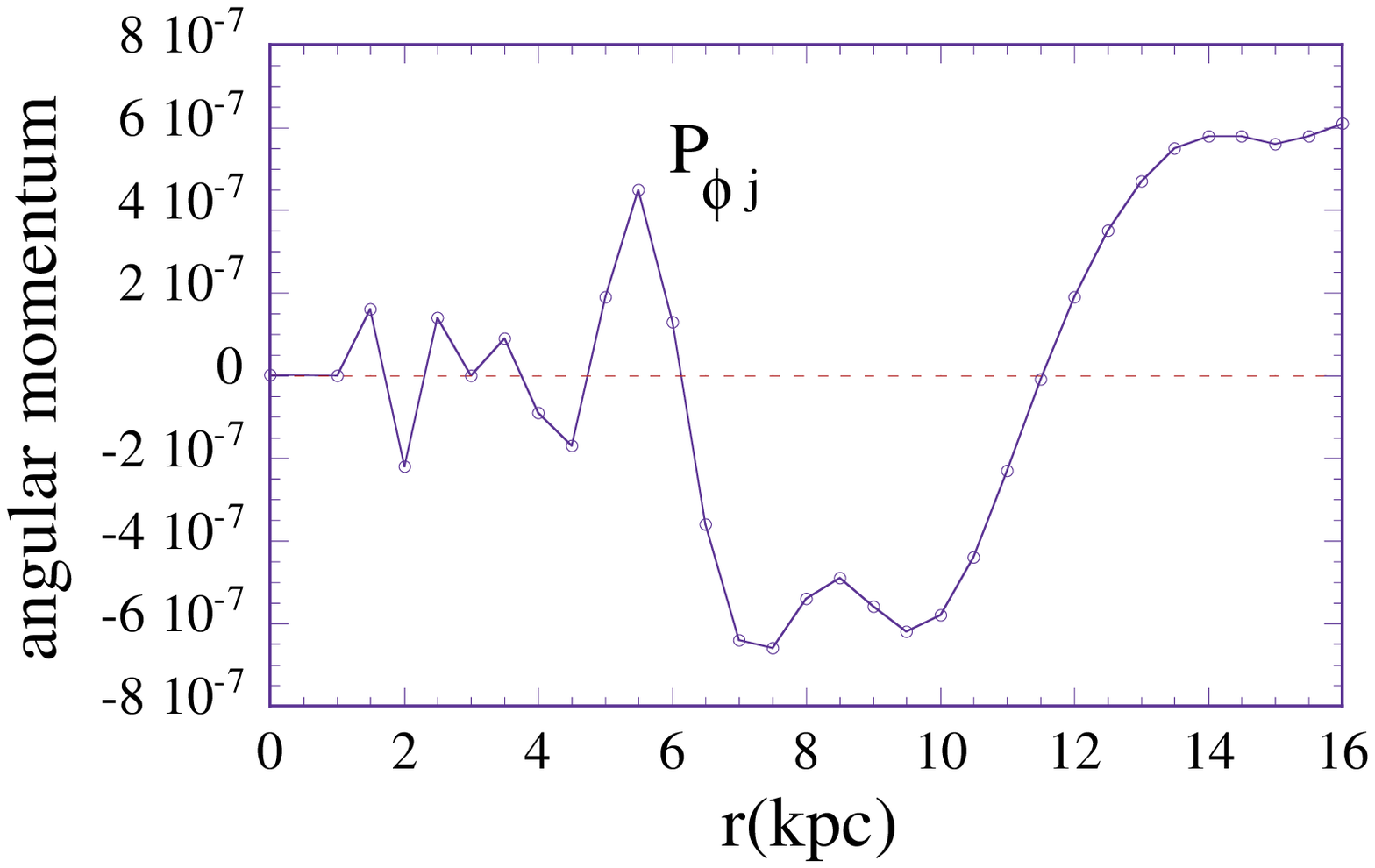}
\caption{
  Radial dependence of
the perturbations in the ring
angular momenta $P_j$ 
at $t=300$ Myr from
equation (58).  The angular
momentum of the center $P_0$ 
is  negligible.
}
\end{figure*}
%%%%%%%%%%%%%%%%%%%%%%%%%%%%%%%%%%%%%%%%%%

%%%%%%%%%%%%%%%%%%%%%%%%%%%%%%%%%%%%%%%%%%
\begin{figure*}[b]
\epsfscale=500
\plotone{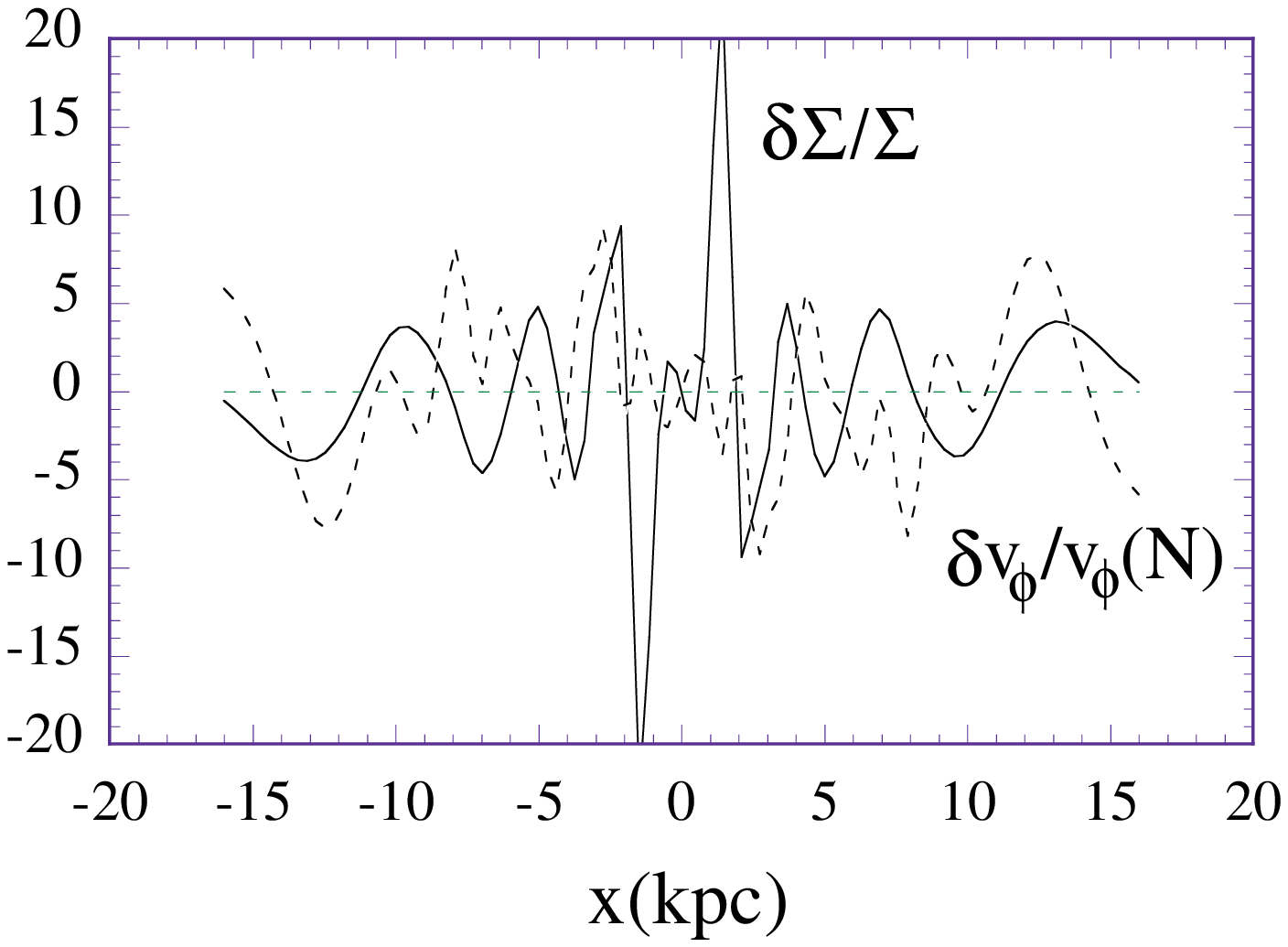}
\caption{
Profiles of fractional
density variations $10^2\delta \Sigma(x,0)
/\Sigma(x,0)$ (in percent) and variation of 
azimuthal velocity 
$10^3\delta v_y(x,0)/v_\phi(N)$
along the $x-$axis through the
middle of the galaxy at $t=400$ 
Myr for the same
case as Figure 10. 
   Here, $v_\phi(N)\approx 265$ km/s
is the disk rotation velocity
at $r=16$ kpc. 
   The vertical scale is
 arbitrary in that
the equations solved are linear,
but the ratio 
$\delta \Sigma/\delta v_\phi$ 
is fixed.
}
\end{figure*}
%%%%%%%%%%%%%%%%%%%%%%%%%%%%%%%%%%%%%%%%%%

    Figure 10 shows the essential
behavior in a polar plot of the shifts
and displacements at two times.
  The shift of the center is
negligible on the scale of the
plot.
  The magnitude of azimuthal
displacement of the outer ring
$\delta_N$ exhibits
an approximately {\it linear} growth
with time, 
$\delta_N \approx {\rm const}+t/660{\rm Myr}$,
  for the considered initial conditions
and $t\lesssim 1$ Gyr.
  In contrast, the magnitude of
the radial shifts $\epsilon_j$
remains bounded by its initial
maximum value $\epsilon_N(t=0)$.

%%%%%%%%%%%%%%%%%%%%%%%%%%%%%%%%%%%%%%%%%%
\begin{figure*}[t]
\epsfscale=500
\plotone{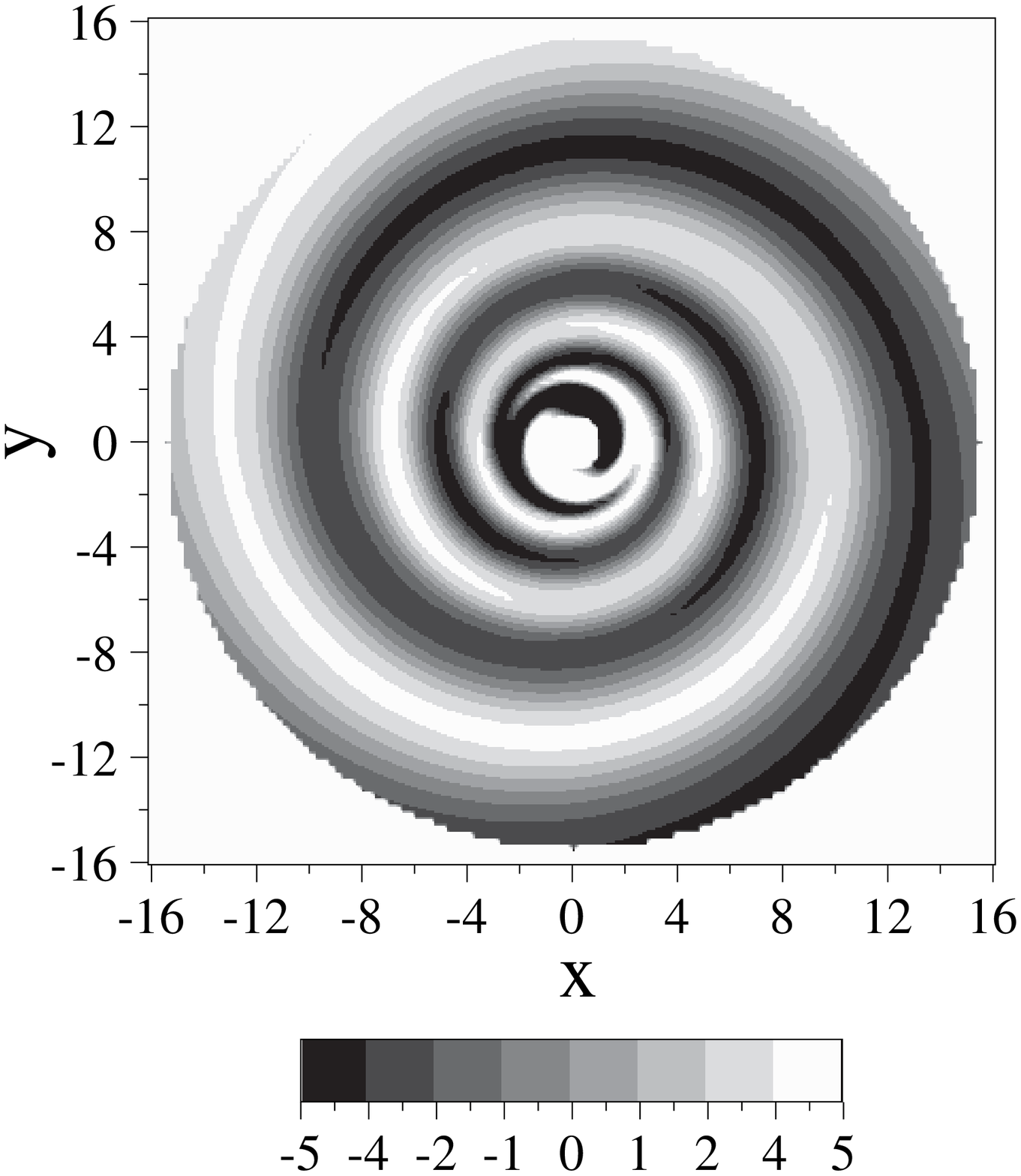}
\caption{
Two-dimensional
appearance of the fractional
density variations $10^2 \delta \Sigma(x,y)
/\Sigma(x,y)$ at $t=400$ Myr for the
same case as Figure 13. 
}
\end{figure*}
%%%%%%%%%%%%%%%%%%%%%%%%%%%%%%%%%%%%%%%%%%

  The  patterns formed
by both $\delta_j$ and $\epsilon_j$
in Figure 10
are {\it trailing} spirals.
  This is different from the proposal
of Baldwin {\it et al.} (1980) that
leading spirals should form.
  The rotation of the outer point
on the spiral $\Delta_N$($r_N=16$ kpc)
in Figure 10 is in the direction
of rotation of the disk matter.
Its pattern speed $\Omega_p$ corresponds
to a  period
 $2\pi/\Omega_p\approx 460$ Myr, which is
a factor $\approx 1.24$
longer than the rotation period of matter at
this radius ($\approx 370$ Myr). 
  The pattern period of  
say $\Delta_{15}(r=8{\rm kpc})$ is
$\approx 188$ Myr, which is a
less than the rotation period of the matter
at this radius ($\approx 206$ Myr).
  Thus, it is evident that the
spiral is ``wrapping up'' as
time increases.
   At the same time, the spiral
pattern propagates radially
outward.  The outward speed is
about $10$ km/s at $r\sim 8$ kpc for 
$t\sim 300$ Myr.

  Figure 11 shows a polar plot 
of the radius
to the maximum of the azimuthal
displacement $r_j(\psi_j)$ at
two times.
 The curve is a  trailing
spiral with
an approximate fit given by 
\begin{equation}
\psi =
A(t)
\exp\left(- {r\over a(t)}
\right)~,
\end{equation}
where $A \approx 0.065$ 
$(t/{\rm kpc})$ rad, and
$a \approx 7.0+0.0044 $
$ (t/{\rm Myr})$ kpc
for $t\lesssim 1$ Gyr.
  Thus the radial spacing between
spiral arms is 
$\lambda_r \approx
(2\pi a/A)\exp(r/a)$ for $\lambda_r \ll a$.
 For validity of the ring representation
we must have $\lambda_r \geq 2 \delta r(=1$ kpc),
where $\delta r$ the separation between
rings.

  Figure 12 shows the radial variation
of the perturbations in
ring angular momenta.  The perturbation
of the angular momentum of
the  central mass is negligible.
  This figure should be compared with
Figure 8 which gives the same plot
for an exponential disk.

     Some simplification of 
equations is possible in the
present case at long times due to the fact that
$\delta_j\gg \epsilon_j$. 
   Firstly, we have
$\delta{\bf v} \approx \delta v_\phi
\hat{\rvecphi~}$ with
\begin{equation}
\delta v_\phi(r,\phi)
 \approx -(\dot{\delta}_x +\Omega \delta_y)
\sin\phi +(\dot{\delta}_y-\Omega\delta_x)\cos\phi~.
\end{equation}
  Secondly, 
\begin{equation}
 {\delta \Sigma(r,\phi) \over \Sigma(r)}
 \approx{1\over r}
(\delta_x\cos\phi + \delta_y \sin\phi)~,
\end{equation}
for $\delta_j \gg r\epsilon_j/r_d$.

  Figure 13 shows the profiles along
the $x-$axis through the
galaxy center of  the fractional change
in the surface density
$\delta \Sigma/\Sigma$ and
the change in the azimuthal
velocity $\delta v_\phi$ obtained
from (74) and (75).  
   The opposite signs of $\delta v_\phi$
on the two sides of the galaxy would
of course make the rotation curves on
the two sides different as observed
in some cases (Swaters {\it et al.} 1998).
 Note that in some regions the changes
$\delta \Sigma$ and $\delta v_\phi$
are correlated and in other regions
they are anticorrelated.
  Figure 14 shows two-dimensional 
appearance of the fractional surface
density variations from (75).

  For long times $t>1$ Gyr, the
the azimuthal
displacements and shifts of the
inner rings ($2$ and $3$) start 
to become large compared with
the values in the outer disk ($r>4$ kpc)
even though these rings have very
small masses.  
   At the same time,
the displacement of the center,
which has mass $M_0=1.06 \times 10^9M_\odot$,
grows, and at $t=1$ Gyr it is
$\epsilon_0 \approx 0.057$ for the 
conditions of Figures 9-13.  
  If the mass of the center
is $M_0=10^6M_\odot$, then the displacements
and shifts of rings $2$ and $3$ at $t=1$ Gyr
are significantly reduced as is the shift
of the center which is 
$\epsilon_0 \approx 0.014$.

\section{Conclusions}

   The paper develops a theory of
eccentric ($m=\pm1$) linearized
perturbations of an 
axisymmetric disk galaxy 
residing in a spherical
dark matter halo and with a spherical
bulge component. 
    The disk is represented by
a large but finite
number $N$ of rings with shifted
centers {\it and} with perturbed
azimuthal matter distributions.
  This description is appropriate
for a disk with small `thermal'
velocity spread $v_{th}$ where the
matter is in approximately laminar
circular motion.  
   The  spread for a 
thin disk has $(v_{th}/v_\phi)^2 \ll 1$,
but it is sufficent to give
a Toomre $Q(r) \gtrsim 1$. 
  Earlier, Baldwin {\it et al.} (1980)
discussed asymmetries in disk galaxies in terms
of shifted rings but without interactions
between the rings and without azimuthal
displacements of the ring matter.
    Account is taken of the shift 
of the matter at the
galaxy's center, which may include
a massive black hole. 
    The gravitational interactions
between the rings and between the
rings and the center is fully
accounted for, but the halo and
bulge components are treated as
passive gravitational field sources.  
   Equations of motion are derived for
the ring and the center, and from
these we obtain the Lagrangian for
the rings$+$center system.
   For this system 
we derive an energy 
constant of the motion, 
and  a total canonical angular
momentum constant of the motion.

    We first  discuss the nature
of the precession of
a single ring with the other
rings fixed;  this case although
 not self-consistent is informative.
    There are four modes,  analogs to
the normal modes of a non-rotating
system, and  two  have negative energy
and two positive energy.
  Negative energy modes are unstable
in the presence of dissipation such
as that due to dynamical friction.
  We go on to study the eccentric motion
of a disk consisting 
of two rings of different radii
but equal mass $M_d/2$.  
  Above a threshold value of $M_d$ the
two rings are unstable with
instability
due  merging of positive
and negative energy modes. 
  This result is obtained by solving
the eighth order polynomial for the
frequencies of the eight modes. 
  Above a second, somewhat larger threshold
value of $M_d$, a second instability
appears, and in this case the ring
motion is such that the angular
momentum of the inner ring decreases
while that of the outer ring increases.
  For the unstable motion, the maximum
of the azimuthal density enhancement
of a ring occurs at an angle about $180^\circ$
from the direction of the radial shift.
This allows the center of mass of the ring
to move closer to the center of mass of
the other ring and to the origin.

  We also analyze the eccentric motion of
a disk of one ring interacting 
with a radially shifted central mass.
 This system  has six modes, the frequencies
of which are obtained by solving a sixth
order polynomial. 
    In this case, instability sets
in above a threshold value of the central
mass (for a fixed ring mass), and it  
acts to increase the angular momentum of
the central mass (which therefore rotates
in the direction of the disk matter), while
decreasing the angular momentum of the ring.
  The instability is again due to the merging
of positive and negative energy modes.

   We study the eccentric dynamics
of a disk with an exponential surface density
distribution represented by a large number
$N=31$ of rings and a central 
mass $M_0 \sim 10^9M_\odot$
which may include the mass of a  black hole.
  The outer radius of the 
disk is $r_N =16$ kpc;
  we have checked that this value has 
negligible affect
on the reported results.
   In this case, we numerically integrate
the equations of motion.
  A check on the validity of the integrations
is provided by monitoring the mentioned
total energy and total canonical angular
momentum, which are found to be accurately
constant in all presented results.
    The inner part of the disk 
$r\lesssim 2.5$ kpc is
found to be strongly unstable
with $e-$folding time $\sim 30$ Myr for the
conditions considered.  The $e-$folding time
is somewhat longer if $M_0=0$.
  Angular momentum of the rings is
transferred {\it outward}, {\it and} to
the central mass if it is  present. 
  A {\it trailing} one-armed spiral 
wave is formed in the disk. 
  This differs from the prediction of
Baldwin {\it et al.} (1980) of a 
leading one-armed spiral.
   The outer part of the disk $r \gtrsim r_d$
is stable and in this 
region the angular momentum is
transported by the wave. 
  Thus our results appear compatible
with the theorem of Goldreich and 
Nicholson (1989) regarding
angular momentum in {\it stable} rotating
fluids. 
   The   instability found here
appears qualitatively 
similar to that found
by Taga and Iye (1998b) for a fluid
Kuzmin disk with surface density
$\Sigma \propto 1/(1+r^2)^{3/2}$
with a point mass at the center where
unstable trailing one-armed spiral
waves are found. 

   The present
linear theory does not address
the issue of saturation of 
growth of the eccentric motion.
   One possibility is that 
the strong  instability
of the inner rings of the disk
leads to the destruction of this
part of the disk.
  For this reason 
we have studied a disk 
with a modified exponential
density distribution where 
the surface density of
the inner part of the disk is reduced.
   However, the mass of the center of
the galaxy was kept the same as in
the case of an exponential disk,
$M_0 \sim 10^9 M_\odot$. 
  In this case we find much slower, linear -
as opposed to exponential -
growth of the eccentric motion of 
the disk for times $t \lesssim 1$ Gyr. 
 A  trailing one-armed  spiral wave 
forms in the disk and  becomes more
tightly wrapped as time increases.
  Angular momentum is transferred outward.
  The motion of the central mass if present 
is small compared with that of the disk
for $t \lesssim 1$ Gyr.

  For long times $t>1$ Gyr, the
the azimuthal
displacements and shifts of the
inner rings  start 
to become large compared with
the values in the outer disk.
   At the same time,
the radial shift of the center
grows.  
  This shift is significantly
reduced if the mass of the center
is changed from $\sim 10^9$ 
to $10^6 M_\odot$.

\acknowledgments{We thank M.S.  Roberts
for valuable discussions which
stimulated this work.  
    We thank R.H. Miller
for bringing the work of Taga and Iye to
our attention and for
valuable comments on  our work.
  We thank M.M. Romanova for
valuable discussions and
an anonymous referee for valuable 
comments.
  This research has
been partially supported by NSF
grant AST 95-28860 to M.P.H. and
NASA grant NAG5 6311 to R.V.E.L.}

\onecolumn

\appendix
\section{{Tidal Coefficients}}

  For $|r_j-r_k| \gg \sqrt{\Delta r_j\Delta r_k}$,
the ring profiles can be treated as delta functions,
$S(r|r_j) \rightarrow \delta(r-r_j)/r$,
and consequently the `tidal coefficients' 
of equations (29) - (32) 
can be simplified to give
\begin{eqnarray}
C_{jk} &\approx& GM_jM_k~ 
{\partial^2 {\cal K}(r_j,r_k) 
\over \partial r_j ~\partial r_k}= -~ {2G M_jM_k 
\over \pi (r_j+r_k)(r_j-r_k)^2}~E(k_{jk})~,\\
D_{jk}&\approx&
GM_jM_k~
{\partial {\cal K}(r_j,r_k)
\over r_k ~\partial r_j}
= -~{GM_jM_k\over \pi r_j^2 (r_j^2-r_k^2)}
\bigg[(r_j+r_k)E(k_{jk})+(r_j-r_k)K(k_{jk})\bigg]~,\\
D^{\prime}_{jk}& \approx&
GM_jM_k~
{\partial {\cal K}(r_j,r_k)
\over r_j ~\partial r_k}=
 D_{kj}~,\\
E_{jk}& \approx& 
GM_jM_k~{{\cal K}(r_j,r_k) \over r_j ~r_k}=
{GM_jM_k\over \pi r_j^2 r_k^2 (r_j+r_k)}
\bigg[(r_j^2+r_k^2)K(k_{jk})-(r_j+r_k)^2 E(k_{jk})
 \bigg]~,
\end{eqnarray}
(Mathematica V. 3) where $k_{jk} 
\equiv 2 \sqrt{ r_j r_k}/(r_j+r_k)$,
where ${\cal K}$ is defined by equation (31), 
and where 
$$
K(k)\equiv \int_0^{\pi/2} d\phi 
\bigg/\sqrt{1-k^2\sin^2\phi}
~,~~{\rm and}~~E(k)\equiv \int_0^{\pi/2} d\phi
\sqrt{1-k^2\sin^2\phi}~,
$$
are complete elliptic integrals of the 
first and second kinds respectively.
Note that for $r_k \gg r_j$, 
  $C_{jk} \approx - GM_jM_k/r_k^3$, 
$D_{jk} \approx GM_jM_k/(2r_k^3)$,
$D^\prime_{jk} \approx -GM_jM_k/r_k^3$, 
and $E_{jk} \approx GM_jM_k/(2r_k^3)$.

In the opposite limit where $|r_j-r_k|
\ll \sqrt{r_j r_k}$, equations (27)-(30)
can be evaluated approximately as
\begin{eqnarray}
C_{jk} &\approx& 
{GM_jM_k \over 8\pi\sqrt{\pi}(\Delta r)\bar r^2}
\int_{-\infty}^\infty
{y dy ~\exp(-y^2/4) \over r_k-r_j +y \Delta r}
\bigg\{2\bar r-(r_k-r_j+y\Delta r)
\ln\bigg[{8\bar r \over
|r_k-r_j+y\Delta r|}\bigg] \bigg\}~,\nonumber \\
&\approx&{GM_jM_k \over 2\pi (\Delta r)^2 \bar r}
\bigg[1-{\sqrt{\pi} \over 2} u \exp(-u^2/4) 
{\rm erfi}(u/2) +{\cal O}(\Delta r /\bar r)\bigg]~,
\end{eqnarray}
where $\bar r \equiv (r_j+r_k)/2$, $\Delta r 
\equiv \sqrt{(\Delta r_j^2 +\Delta r_k^2)/2}$, $u \equiv
(r_k-r_j)/\Delta r$, and
and ${\rm erfi}(x)
\equiv{\rm erf}(ix)/i = 
(2/\sqrt{\pi})\int_0^x dy\exp(y^2)$.
The integral in (A5) is a principal value integral
of the form occurring in the plasma dispersion
function W (Ichimaru 1973).  
Also,
\begin{eqnarray}
D_{jk} &\approx& 
{GM_jM_k \over 4\pi\sqrt{\pi}~\bar r^3}
\int_{-\infty}^\infty
{dy ~\exp(-y^2/4) \over r_k-r_j +y \Delta r}
\bigg\{2\bar r - (r_k-r_j+y\Delta r)
\ln\bigg[{8(\bar r/\Delta r) \over
|u+y|}\bigg] \bigg\}~,\nonumber \\
&\approx& -~{GM_jM_k \over 4 \pi \sqrt{\pi}~ \bar r^3}
\bigg\{\int_{-\infty}^\infty dy~ \exp(-y^2/4)
\ln\bigg[{8\bar r /\Delta r \over |u+y| }\bigg]
-2\pi(\bar r/\Delta r)\exp(-u^2/4) {\rm erfi}(u/2) \bigg \}~,
\end{eqnarray}
and
\begin{equation}
E_{jk}  \approx  {GM_jM_k \over 2\pi \sqrt{\pi} \bar r^3}
\int_{-\infty}^\infty dy \exp(-y^2/4)
\ln \bigg[ {1.0827 (\bar r/\Delta r) \over
|u+y|}\bigg]~.
\end{equation}
From these expressions we obtain
\begin{equation}
C_{jj}\approx {G M_j^2  \over 2\pi r_j(\Delta r_j)^2}~,\quad
D_{jj} \approx -~{G M_j^2 \over 2\pi r_j^3} 
\ln \bigg({10.68r_j\over \Delta r_j}\bigg)~, \quad
E_{jj} \approx {G M_j^2 \over \pi r_j^3}
\ln \bigg({1.445 r_j \over \Delta r_j }\bigg) ~.
\end{equation}
Equations (A1) - (A8) are valuable for numerical
evaluation of the tidal coefficients.

In the following we derive some useful relations 
involving the tidal coefficients.
  From equation (12) we have
\begin{equation}
\Omega_d^2(r) ={1\over r}{\partial \Phi_d \over \partial r}
=2\pi G\int_0^\infty r^\prime d r^\prime \Sigma_d(r^\prime)
\oint {d \Psi \over 2\pi} 
{\left(1-r^\prime \cos \Psi /r \right)
\over \left[r^2+(r^\prime)^2-2r r^\prime \cos\Psi \right]^{3/2}}~.
\end{equation}
The $\Psi$ integral in this case is
\begin{equation}
{1 \over \pi r^2 [r^2 - (r^\prime)^2]}
\left[ (r+r^\prime) E + (r-r^\prime)K \right]~.
\end{equation}
Comparison with (A2) shows that 
shows that
%%%%%%%%%% eqn11
\begin{equation}
\Omega_{dj}^2\equiv \int_0^\infty r dr S(r|r_j)\Omega_d^2(r)
= -~{1\over M_j}\sum_{k=1}^N D_{jk}~.
\end{equation}
   This expression does not include the disk mass within
the inner ring.  
  As discussed in \S 2.6, this part of the
disk is treated as a point mass $M_0$ with
unperturbed position ${\bf r}=0$.
   If there is a central black hole $M_{bh}$,
its mass is include in $M_0$.  
   To account for
the influence of $M_0$, we simply 
add the term $GM_0/r_j^3$
to the right hand side of (A11).
    The resulting expression for
$\Omega_{dj}$  is useful for the 
numerical calculations.   
  For  $N \gtrsim 30$ and $r_j =1$ to $10-20$ kpc, we find
that equation (A11) gives accurate agreement
with the analytic expression (3).

 An alternative expression for $\Omega_d^2$
can be obtained by integration by parts,
\begin{equation}
\Omega_d^2(r)=-~{2\pi G \over r}\int_0^\infty
r^\prime dr^\prime {\partial \Sigma^\prime 
\over \partial r^\prime}
~{\cal K}(r,r^\prime)~.
\end{equation}
Thus
$$
M_j\left(r{\partial \Omega_d^2 \over \partial r}\right)_j=
\int r dr S(r|r_j) r{\partial \Omega_d^2(r) \over \partial r} =
GM_j\sum_k M_k\int \!\!\int r dr~r^\prime dr^\prime
S(r|r_j){\partial S(r^\prime|r_k) 
\over \partial r^\prime}
{\partial [ {\cal K}(r,r^\prime)/r] 
\over \partial r}~,
$$
or
\begin{equation}
M_j\left(r{\partial \Omega_d^2 \over \partial r}\right)_j=
GM_j\sum_kM_k \int\!\int rdr~r^\prime dr^\prime S(r|r_j)
\bigg[{1\over r^\prime}{\partial [r^\prime S(r^\prime|r_k)] \over \partial r^\prime}
-{S(r^\prime|r_k) \over r^\prime}\bigg]
\bigg[ {{\cal K}(r,r^\prime) \over r} - 
{\partial {\cal K}(r,r^\prime) \over \partial r}
\bigg]
\end{equation}
In view of equations (27) - (30), this equation can be written 
as
\begin{equation}
M_j\left(r{\partial \Omega_d^2 \over \partial r}\right)_j=
\sum_k(C_{jk}+D_{jk}) - \sum_k(D_{jk}^\prime +E_{jk})~.
\end{equation}
This relation is useful in \S2.5.

We also evaluate the disk gravitational
potential in the ring representation,
\begin{eqnarray}
\Phi_d(r)&=&-G\int d^2r^\prime~ {\Sigma_d(r^\prime)
\over |{\bf r}-{\bf r}^\prime|} 
= -~2\pi G \int_0^\infty r^\prime dr^\prime~\Sigma_d(r^\prime)
\oint{d \Psi \over 2\pi} 
{1 \over [r^2+(r^\prime)^2 -2 rr^\prime \cos\Psi]^{1/2}} \nonumber\\
&=& -2\pi G\int r^\prime dr^\prime~ \Sigma_d(r^\prime)
{2 K(k) \over \pi (r+r^\prime)}~.
\end{eqnarray}
With
\begin{equation}
\Phi_{dj} \equiv  \int_0^\infty 
rdr~S(r|r_j)\Phi_d(r) 
= {1\over M_j}\sum_k \Lambda_{jk}~,
\end{equation}
where 
\begin{equation}
\Lambda_{jk} =-GM_jM_k\int\int rdr~r^\prime dr^\prime 
~S(r|r_j)S(r^\prime|r_k)
~{2 K(k) \over \pi(r+r^\prime)}
~.
\end{equation}
For $|r_j-r_k| \gg \sqrt{\Delta r_j \Delta r_k}$, 
\begin{equation}
\Lambda_{jk} \approx -GM_jM_k{2 K(k) \over \pi (r_j+r_k)}~,
\end{equation}
whereas for $|r_j-r_k| \ll \sqrt{r_jr_k}$
\begin{equation}
\Lambda_{jk} \approx -GM_jM_k{1\over 2\pi \sqrt{\pi} ~\bar r}
\int_{-\infty}^\infty dy \exp(-y^2/4) \ln\bigg[ 
{8\bar r /\Delta r \over |u+y| }\bigg]~,
\end{equation}
where $u \equiv (r_k-r_j)/\Delta r$ as above.
Note that $\Lambda_{jk}=\Lambda_{kj}$ and
that $\Lambda_{jj} \approx 
-GM_j^2(1/\pi r_j)\ln(10.68 r_j/\Delta r_j)$.

\end{document}